\documentclass[aps,prmaterials,reprint,superscriptaddress,amsmath,amssymb,
               floatfix,longbibliography]{revtex4-2}

\usepackage{graphicx}
\usepackage[T1]{fontenc}
\usepackage{bm}
\usepackage{xcolor}
\usepackage[colorlinks=true,allcolors=blue]{hyperref}

\begin{document}

\title{Programming the Nonlinear Optics\\ of a Nanoporous Photonic Metamaterial by Confined Crystallization}

\author{Andrii Z. Maxym}
\affiliation{Lviv National Polytechnic University, 12 Bandery str., 79013 Lviv, Ukraine}
\author{Junhui Lang}
\affiliation{University of Angers, LPhIA, SFR MATRIX, 2 Bd. Lavoisier, Cedex 01, 49045 Angers, France}
\author{Bouchta Sahraoui}
\affiliation{University of Angers, LPhIA, SFR MATRIX, 2 Bd. Lavoisier, Cedex 01, 49045 Angers, France}
\author{Anatoliy S. Andrushchak}
\affiliation{Lviv National Polytechnic University, 12 Bandery str., 79013 Lviv, Ukraine}
\author{Yaroslav Shchur}
\affiliation{Yukhnovskii Institute for Condensed Matter Physics of NASU, 1 Svientsitskii str, Lviv 79011, Ukraine}
\author{Olha Kityk}
\affiliation{Energia Oze Sp. z o.o., ul. Cz{\k{e}}stochowska 7, 42-274 Konopiska, Poland}
\author{Robert Wielgosz}
\affiliation{Energia Oze Sp. z o.o., ul. Cz{\k{e}}stochowska 7, 42-274 Konopiska, Poland}
\author{Monika Lelonek}
\affiliation{SmartMembranes GmbH, Heinrich-Damerow-Str. 4, 06120 Halle (Saale), Germany}
\author{Piotr Pawlik}
\affiliation{SmartMembranes GmbH, Heinrich-Damerow-Str. 4, 06120 Halle (Saale), Germany}
\author{Patrick Huber}
\email{patrick.huber@tuhh.de}
\affiliation{Institute for Materials and X-ray Physics, Hamburg University of Technology, Denickestr. 15, 21073 Hamburg, Germany}
\affiliation{Centre for X-ray and Nano Science CXNS, Deutsches Elektronen-SynchrotronDESY, Notkestr. 85, 22607 Hamburg, Germany}
\author{Jaros{\l}aw J{\k{e}}dryka}
\affiliation{Faculty of Electrical Engineering, Cz{\k{e}}stochowa University of Technology, Al. Armii Krajowej 17, 42-200 Cz{\k{e}}stochowa, Poland}
\author{Andriy V. Kityk}
\email{andriy.kityk@pcz.pl}
\affiliation{Faculty of Electrical Engineering, Cz{\k{e}}stochowa University of Technology, Al. Armii Krajowej 17, 42-200 Cz{\k{e}}stochowa, Poland}

\date{\today}

\begin{abstract}
A membrane traversed by a dense array of nanochannels acts as a photonic metamaterial: with channels far smaller than the wavelength of light, the optical response is set by the architecture rather than by the constituent materials. Here each nanochannel is turned into a nonlinear optical meta-atom by filling anodic aluminum oxide (AAO) with benzil (BZL), a noncentrosymmetric molecular crystal, and the crystallization pathway programs their orientation. In free-standing membranes every channel solidifies independently, the orientations remain uncorrelated, and the effective second-harmonic generation (SHG) response is nearly polarization isotropic; its magnitude follows a core--shell model in which an optically inactive interfacial layer of 2.6 nm reduces the active volume of each meta-atom, reproducing on bare alumina the value previously found on alkyl-grafted silica and indicating a host-insensitive limit. Seeding crystallization from a bulk benzil layer instead aligns the meta-atoms over submillimeter domains and produces a macroscopically anisotropic effective susceptibility. Channel diameters from 32 to 270 nm span the transition out of the subwavelength regime, and the anomalous directional SHG emission reported here is confined to the metamaterial limit. Confined crystallization thus offers a route to nonlinear optical metamaterials whose effective susceptibility tensor is written during processing.
\end{abstract}

\keywords{Photonic metamaterials, Meta-atoms, Nanoconfinement, Second harmonic generation, Nanocomposites, Molecular crystals}

\maketitle

\section{Introduction}\label{sec:intro}

Nonlinear optical (NLO) materials are of fundamental and technological interest for optical frequency conversion, laser technology, optical modulation, photonic signal processing, and emerging integrated photonic devices~\cite{Kang2022,Zyss1994b,Prasad1991,Mutailipu2025}. Among the various nonlinear processes, second-harmonic generation (SHG) is particularly sensitive to the symmetry and structural organization of the nonlinear-optically active medium. In molecular and crystalline materials, the macroscopic second-order NLO response is determined not only by the molecular hyperpolarizability but also by crystal symmetry, molecular packing, and the spatial and orientational organization of the NLO-active units~\cite{Kang2022,Zyss1994b,Prasad1991,Zyss1993}. Consequently, controlling the structural organization of NLO materials provides an important route toward tailoring their optical response. The NLO response can be further engineered through structural modification, chemical substitution, or confinement-induced changes in the local crystal environment~\cite{He2023c}. Nanostructuring provides another route for controlling NLO interactions by modifying light--matter coupling and the electromagnetic environment~\cite{Talts2023s}. Recent studies of molecular and low-dimensional crystals have further demonstrated that structural dissymmetry, crystallographic orientation, and the arrangement of nonlinear-active units can strongly affect both the magnitude and polarization dependence of SHG~\cite{Zyss1993,Fang2023,Zhou2024n,Xu2024f}.

Organic molecular crystals are particularly attractive for NLO applications because their optical properties can be modified through molecular structure, intermolecular interactions, and crystal packing~\cite{Zyss1994b,Prasad1991}. Engineering the molecular orientation in such materials can yield large macroscopic second-order responses~\cite{Jeong2011}, and dedicated crystal-growth strategies have been developed to obtain organic single crystals with large second-order susceptibilities~\cite{Yang2005}. Benzil (BZL, (C$_6$H$_5$CO)$_2$) represents a convenient model molecular crystal for investigating such structure--property relationships. BZL crystallizes in a noncentrosymmetric trigonal structure with the $P3_{1}21$ space group~\cite{Brown1965,More1987}, allowing second-order nonlinear optical activity~\cite{Elkarout2023}. Its NLO properties include measurable SHG~\cite{Elkarout2023} as well as pronounced third-order effects, including saturation absorption~\cite{Saranraj2018,Kanagasekaran2008} and a negative nonlinear refractive index~\cite{Elbassuony2023i,Saranraj2018,Kanagasekaran2008,Rajalakshmi2012}. BZL also exhibits a relatively high laser-damage threshold~\cite{Elbassuony2023i,Saranraj2018} and sufficient optical transparency in the visible spectral range, with an experimentally reported absorption cutoff around 430--440~nm~\cite{Saranraj2018,Kanagasekaran2008}. This transparency enables transmission of the second-harmonic radiation generated under 1064~nm excitation. Importantly for nanocomposite fabrication, BZL has a relatively low melting point of approximately 95~$^\circ$C~\cite{Acree2017}, enabling its infiltration into porous hosts from the melt followed by crystallization upon cooling~\cite{Elkarout2023}. These characteristics make BZL particularly suitable for studying how the confinement and crystallization of an organic molecular crystal influence its macroscopic nonlinear optical response.

Spatial confinement in nanoporous hosts provides an additional structural parameter for controlling molecular crystallization. When a molecular material is confined within nanoscale channels, the competition between guest--host interactions, surface effects, nucleation, crystal growth, and geometrical restrictions can modify the dimensions, morphology, orientation, and phase behavior of the resulting crystalline phase~\cite{Alba-Simionesco2006, Huber2015}. More generally, nanoscale confinement can alter nucleation and crystal-growth mechanisms, phase stability, and polymorphism~\cite{Duran2011, Suzuki2013, Suzuki2015, Henschel2010PolymorphismSilicon, Hillmyer2012}. Such effects are especially relevant to nonlinear optics because the macroscopic SHG response depends on the collective contribution of the nonlinear-active domains and is therefore sensitive to their spatial and crystallographic organization~\cite{Zyss1993,Fang2023,Zhou2024n}. Nanoporous matrices can thus act not merely as mechanically supporting phases but as structural environments capable of modifying and engineering the optical functionality of confined molecular materials, in line with the wider development of organic--inorganic hybrid materials for optoelectronic applications~\cite{Zhou2024}.

Our previous study demonstrated pronounced SHG from BZL nanocrystals confined within functionalized mesoporous silica channels with diameters of approximately 6--13~nm~\cite{Elkarout2023}. Capillary crystallization produced oriented nanocrystalline BZL aggregates, while X-ray diffraction revealed a pronounced crystallographic texture whose relative contributions depended on the channel diameter. The SHG response exhibited a strong dependence on the degree of spatial confinement, which was interpreted in terms of a heterogeneous morphology comprising a disordered, SHG-inactive interfacial region adjacent to the pore wall and a crystalline core contributing predominantly to the SHG response~\cite{Elkarout2023}. This interpretation suggested that the fraction of the nonlinear-optically active crystalline phase increases with increasing channel diameter. However, the relatively narrow range of pore dimensions accessible in the silica host limited the possibility of testing this confinement effect over a substantially broader length scale and did not allow the influence of the crystallization pathway to be separated from the effect of extreme spatial confinement.

Anodic aluminum oxide (AAO) membranes provide an attractive platform for extending this investigation. Electrochemical anodization of aluminum produces porous alumina membranes containing dense arrays of approximately cylindrical nanochannels, whose dimensions and ordering can be controlled through the anodization conditions~\cite{Furneaux1989f,Masuda1995aao,Li1998aao}. The development of controlled-porosity anodic aluminum oxide established AAO as a versatile membrane and nanostructuring platform~\cite{Furneaux1989f}, while subsequent advances in anodization enabled highly ordered, long-range and high-aspect-ratio pore arrays with improved structural uniformity~\cite{Lee2006aao}. The nearly parallel channel architecture, mechanical robustness, and broad range of accessible pore dimensions make AAO particularly suitable for studying confined crystallization and guest--host interactions~\cite{Thompson1997p,Huber2015}. In addition, the unidirectional geometry of AAO introduces a well-defined macroscopic structural axis that can couple the organization of the confined guest phase to its optical properties.

Beyond providing a static geometric constraint, nanoporous hosts can also control the orientation of a confined crystalline phase. Confinement can moreover select the crystallization pathway itself, as in situ studies of nucleation and polymorph selection in nanoscale pores have shown~\cite{Anduix2021}. Steinhart et al. showed that polymers crystallized inside AAO nanochannels form macroscopic arrays with uniform crystal orientation when a bulk-like surface film connects the individual nanostructures, the dominant growth direction then being aligned with the long axes of the template pores~\cite{Steinhart2006}; the confined crystallization of polymers in AAO templates has since been reviewed in detail~\cite{Michell2014}. The same templating principle has been used to align conjugated polymer chains within AAO nanochannels and thereby imprint a macroscopic optical anisotropy on the composite~\cite{Lo2011}. More broadly, AAO has developed into a versatile platform for three-dimensional nanoarchitectures and functional metamaterials~\cite{Platschek2011,MartinGonzalez2025}. Whether the same principle governs a small-molecule, NLO-active guest, and whether it translates into a measurable macroscopic second-order optical anisotropy, has not been established.

The possibility of using nanoporous alumina and silica to tailor optical anisotropy has been demonstrated particularly clearly for liquid-crystal (LC) nanocomposites. In these systems, the macroscopic optical response results from the interplay between the anisotropic geometry of the host nanochannels, surface anchoring, and the orientational self-assembly of the confined molecules~\cite{Chahine2010,Sentker2019,Kityk2021d}. Studies of LC-filled AAO and porous silica have shown that confinement can modify molecular ordering~\cite{Calus2016c}, phase behavior~\cite{Calus2014,Calus2012}, birefringence~\cite{Chahine2010,Chahine2010CriticalityDisorder}, and electro-optical response~\cite{Busch2017,Kityk2021d}; in the most pronounced cases the channel diameter does not merely perturb the bulk structure but selects discrete, commensurate self-assembled states~\cite{Sentker2018}, while surface functionalization provides an additional means of controlling molecular orientation~\cite{Calus2016c}. In particular, ordered AAO nanochannels have been used as a structural environment for studying and tailoring LC optical anisotropy, including systems in which modification of the pore-wall chemistry changes the resulting molecular organization and effective birefringence~\cite{Calus2016c}. More recently, we demonstrated the confinement-mediated self-assembly of the bent-core nematic CB7CB in AAO and silica membranes with nanochannel diameters extending from a few nanometers to several hundred nanometers~\cite{Maksym2025}. The resulting changes in molecular organization and phase behavior were accompanied by systematic modifications of optical birefringence, demonstrating that geometric confinement can be used as a structural parameter for tailoring the optical response of anisotropic molecular materials~\cite{Maksym2025}. These studies establish nanoporous alumina and silica as versatile platforms for engineering optical anisotropy through confinement and interfacial interactions. However, the corresponding relationship between spatial confinement, crystallization pathway, and the nonlinear optical response of an embedded molecular crystal remains largely unexplored.

This line of work also places the present system in a specific optical context. Photonic metamaterials derive their optical functionality from artificial subwavelength building blocks, so-called \textquotedblleft meta-atoms\textquotedblright, whose geometry and arrangement rather than the properties of the constituent materials define the effective permittivity tensor, giving access to negative refraction, magnetic response at optical frequencies, and engineered anisotropy and chirality~\cite{Smith2004,Soukoulis2011,Kadic2019}. Extending this concept to nonlinear optics is particularly rewarding, because the second- and third-order susceptibilities of such a medium are governed not only by the intrinsic nonlinearity of its constituents but also by local-field enhancement and by the symmetry of the meta-atom arrangement~\cite{Lapine2014,Kruk2017,Zeng2018}. The architecture can even generate a second-order response where the constituents supply none, as demonstrated for photonic-crystal slabs assembled from centrosymmetric materials~\cite{Luo2004}. Most realizations rely, however, on metallic or high-index dielectric resonators defined lithographically, which fixes the optical response at the moment of fabrication, so that subsequent tuning relies on an external stimulus acting on the resonators themselves~\cite{Liu2021met}. 

Self-organized nanoporous solids provide a complementary route, since their channel arrays can be filled, emptied, and refilled with functional guests. A monolithic membrane traversed by a parallel array of cylindrical nanochannels whose spacing is much smaller than the wavelength of light is a photonic metamaterial: its optical functionality is not set by the properties of the constituent materials, but by the architecture, with each filled channel acting as a subwavelength \textquotedblleft meta-atom\textquotedblright~\cite{Sentker2019}. Within this picture the form birefringence of the array and the effective refractive indices of the composite can be tuned through pore size, porosity and filling~\cite{Kumeria2012c,Noyan2022b}, and the orientational order of a confined guest supplies an additional, switchable contribution~\cite{Sentker2019,Kityk2021d}. Second-order nonlinearity has been introduced into such hosts through oriented molecular chromophores~\cite{Waszkowska2021} and, in mesoporous silica, through confined molecular crystals~\cite{Elkarout2023}. What has not been shown is that the macroscopic orientation of these nonlinear meta-atoms can be programmed, that is, that one and the same guest in one and the same host can be switched between a macroscopically isotropic and a macroscopically anisotropic second-order response by the choice of crystallization pathway. That is the question pursued here.

In the present work, we investigate benzil--alumina (BZL:AAO) nanocomposites fabricated using AAO membranes with nanochannel diameters ranging from 32 to 270~nm. Two distinct crystallization pathways are considered: uncorrelated crystallization in free-standing AAO membranes, in which the confined BZL crystallites develop without a common macroscopic crystallographic template, and correlated crystallization initiated by a BZL layer formed at the interface between the AAO membrane and a supporting substrate, from which crystallization propagates into the nanochannels. Raman spectroscopy and X-ray diffraction are employed to characterize the molecular and crystallographic organization of the confined BZL phase in the free-standing membranes, while angle- and polarization-resolved SHG measurements, which probe a far smaller sample area, are used to determine the nonlinear optical response of both crystallization regimes in transmission and reflection geometries. By combining structural characterization with nonlinear optical measurements over a broad range of channel dimensions, we demonstrate two complementary routes for engineering the optical functionality of BZL:AAO nanocomposites. Nanoconfinement controls the magnitude of the nonlinear optical response through the relative contribution of the crystalline phase, whereas the crystallization pathway controls its macroscopic anisotropy through the spatial organization of the BZL crystallites.

\section{Results and Discussion}\label{sec:results}

\begin{figure*}[tp]
	\centering
	\includegraphics[height=0.68\textheight]{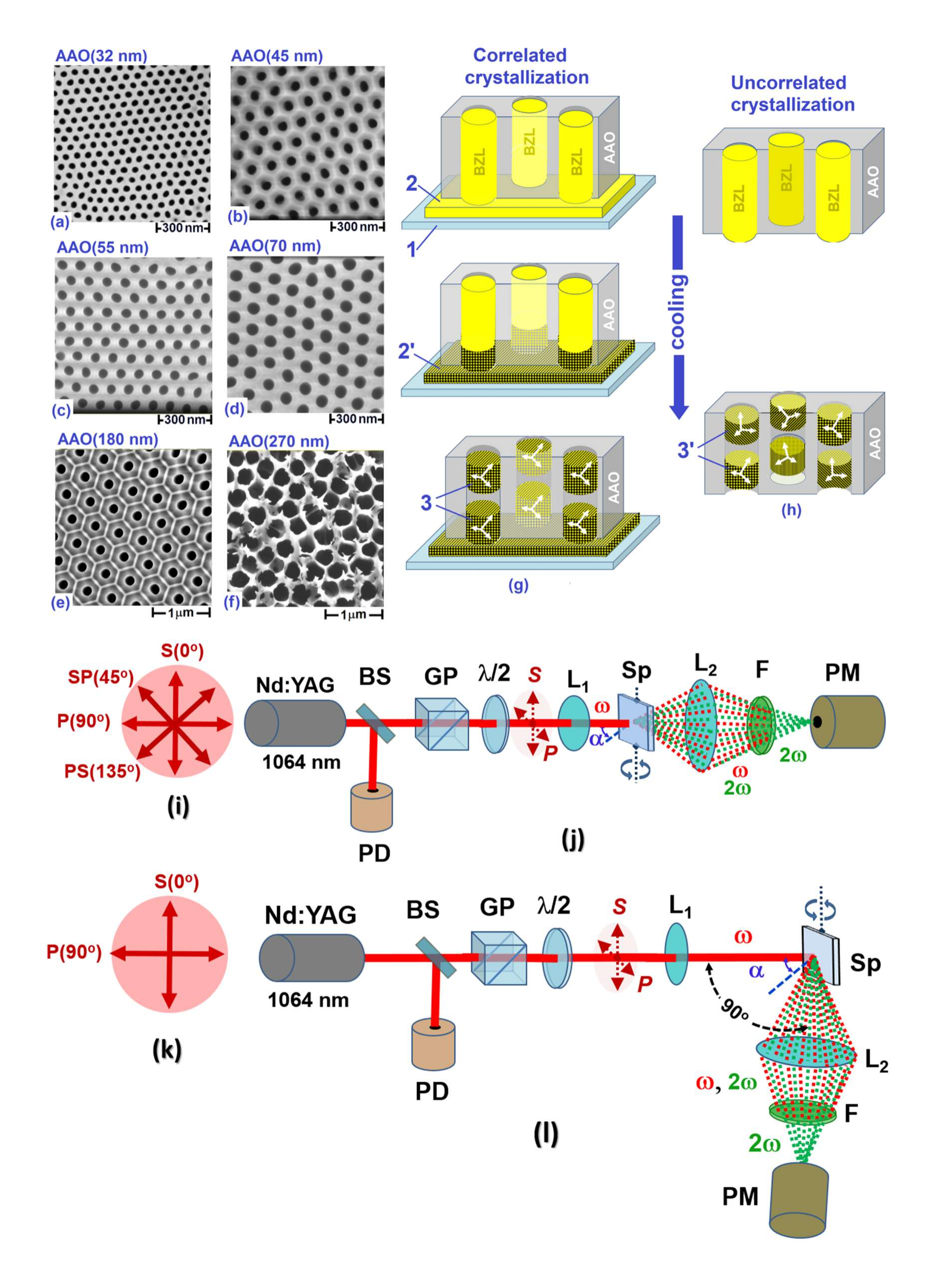}
	\vspace{-0.5cm}
	\caption{Morphology of the AAO templates, schematic representation of the BZL crystallization regimes, and experimental arrangement for SHG measurements. (a--f) SEM images of AAO membranes with average nanochannel diameters of (a) 32, (b) 45, (c) 55, (d) 70, (e) 180, and (f) 270~nm used for fabrication of BZL:AAO nanocomposites. (g,h) Schematic illustration of the two crystallization regimes of BZL confined in AAO nanochannels: (g) correlated crystallization promoted by a bulk BZL layer formed between the AAO membrane and the supporting glass substrate GT;  (h) uncorrelated crystallization in a free-standing AAO membrane. The stages marked 1/1', 2, and 3/3' represent the initial BZL infiltration, crystallization of the bulk interfacial layer followed by crystallization within the nanochannels, and the final state of confined BZL crystallization, respectively. (i) and (k) Definition of the investigated incident linear polarization directions: \textbf{S} (0$^\circ$, vertical), \textbf{SP} (45$^\circ$), \textbf{P} (90$^\circ$, horizontal), and \textbf{PS} (135$^\circ$) employed in SHG measurements in transmission and 90$^\circ$-reflection geometry, respectively. (j) and (l) Schematic diagrams of the experimental setups for azimuthal SHG measurements in transmission and 90$^\circ$-reflection geometry, respectively: Nd:YAG, picosecond laser; BS, beam splitter; GP, Glan--Taylor polarizer; $\lambda/2$, half-wave plate; L$_1$, focusing lens; L$_2$, collection lens; F, interference filter; PM, photomultiplier; PD, photodiode. The sample is rotated by the azimuthal angle $\alpha$ during the measurements.}
	\label{f1}
\end{figure*}

\subsection{Textural Morphology}\label{sec:texture}

The crystallographic texture and structural integrity of the confined BZL phase are central to the nonlinear optical response of the BZL:AAO nanocomposites. In particular, the crystallographic organization established during confinement is expected to determine whether the microscopic nonlinear responses of individual BZL crystallites are averaged out on the macroscopic scale or retained as a measurable optical anisotropy. We first examine the structural state of BZL formed under uncorrelated crystallization conditions in free-standing AAO membranes using Raman spectroscopy and X-ray diffraction.

\begin{figure}[tb]
	\centering
	\includegraphics[width=\columnwidth]{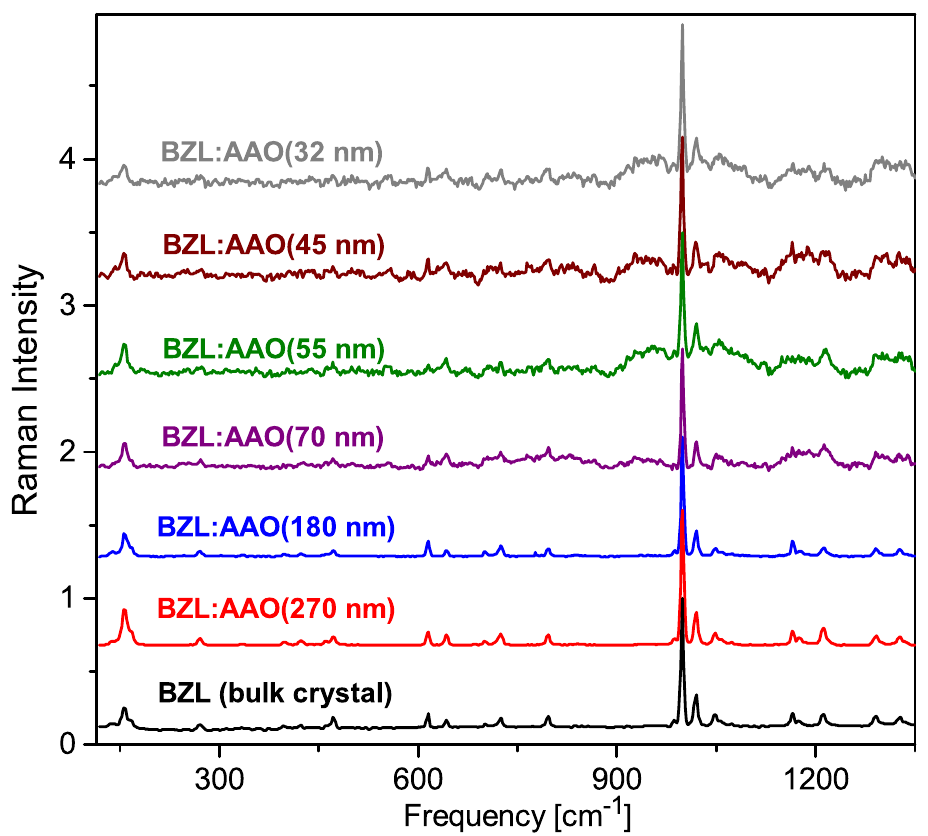}
	\caption{Raman spectra of BZL:AAO nanocomposite membranes with channel diameters of 32, 45, 55, 70, 180, and 270~nm, obtained under free-standing crystallization conditions, with bulk BZL shown for comparison.}
	\label{f2}
\end{figure}

The Raman spectrum of bulk BZL exhibits a characteristic set of vibrational bands, all of which remain recognizable in the BZL:AAO composites, including the prominent bands in the 900--1050~cm$^{-1}$ spectral region (Figure~\ref{f2}). A low-frequency mode centered at approximately 150~cm$^{-1}$ is also observed and can be assigned to an external lattice vibrational mode characteristic of the crystalline phase. The spectral region between 150 and 1500 cm$^{-1}$ contains a series of internal vibrations of phenyl rings and vibrations of C-C and C=O bonds. We carefully investigated the normalised intensity and full width at half maximum (FWHM) of all lines in the Raman spectra taken from BZL:AAO nanocomposite membranes of different diameters and found no distinguishable difference between them. The preservation of the characteristic molecular and lattice vibrational features for all investigated channel diameters demonstrates that confinement within the AAO nanochannels does not disrupt the molecular structure of BZL or prevent its crystallization. These observations are consistent with the XRD results discussed below.

\begin{figure}[tb]
	\centering
	\includegraphics[width=\columnwidth]{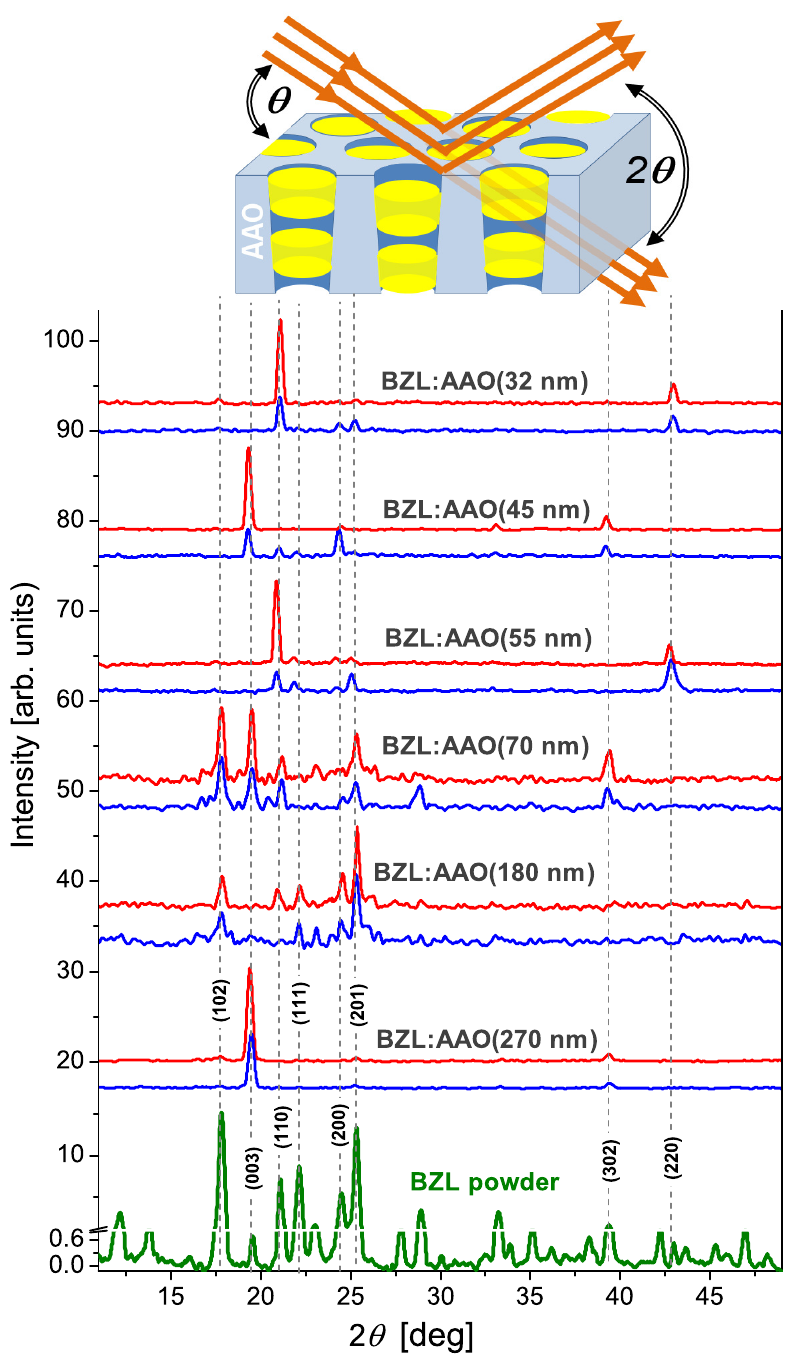}
	\caption{X-ray diffraction patterns ($\theta/2\theta$ scans) recorded from opposite sides of free-standing BZL:AAO nanocomposite membranes with channel diameters of 32, 45, 55, 70, 180, and 270~nm, obtained under free-standing crystallization conditions. The XRD pattern of bulk BZL powder is shown for comparison. The inset schematically illustrates the X-ray diffraction geometry.}
	\label{f3}
\end{figure}

The crystallographic structure and texture of the confined BZL phase were further investigated by X-ray diffraction (XRD) in the $\theta$--$2\theta$ geometry (see the Experimental Section). The elongated geometry of the AAO nanochannels provides a well-defined confinement axis that can influence the nucleation and growth of BZL crystallites and, consequently, their crystallographic texture. The XRD patterns of the BZL:AAO nanocomposites with different nanochannel diameters are presented in Figure~\ref{f3}, together with the bulk BZL pattern for comparison.

For the smallest nanochannels, with diameters of 32, 45, and 55~nm, the diffraction patterns exhibit a pronounced redistribution of the relative intensities of the BZL reflections compared with bulk BZL. The positions of the principal reflections remain consistent with the crystalline BZL phase, whereas their relative intensities are substantially modified. Such a redistribution of the diffraction intensities provides evidence for a modification of the crystallographic texture under strong nanoscale confinement. The confinement therefore affects not only the dimensions of the resulting BZL crystalline regions but also the preferred orientations adopted during nucleation and growth.

With increasing channel diameter, additional BZL reflections become progressively more pronounced, although their relative intensities remain distinct from those of bulk BZL. The patterns obtained for the 70 and 180~nm membranes exhibit a greater number of reflections together with increased peak broadening and structural complexity. These features may reflect a broader distribution of crystallite orientations and/or an increased degree of structural disorder within the confined BZL phase. In contrast, the XRD pattern obtained for the 270~nm membrane contains relatively sharp and intense BZL reflections, suggesting the formation of better-developed crystalline domains within the larger nanochannels.

XRD measurements performed from the opposite side of the membranes reveal the same general BZL reflection sequence but with noticeable differences in relative peak intensities. Such differences may partly originate from variations in the local channel geometry and, consequently, in the confinement conditions along the membrane thickness. Importantly, however, the observed texture does not correspond to a unique crystallographic orientation shared by all BZL crystallites throughout the membrane.

Whereas the lateral size of BZL nanocrystals is constrained by the diameter of the nanochannels, the crystallization front can propagate along the channels over considerably larger distances. By extracting the full width at half maximum (FWHM) and angular position of the XRD peaks from the recorded diffractograms (Figure~\ref{f3}) and applying the Scherrer equation~\cite{Patterson1939}, the average size of the confined BZL nanoclusters along the long channel axis was estimated to be in the range of 22--24~nm. These results suggest that spontaneous, uncorrelated crystallization within the nanochannels involves the formation of liquid-phase capillary bridges separated by voids, which subsequently crystallize into crystalline nanoclusters. In particular, the appearance of such nanoclusters accounts for the pronounced optical scattering observed during the SHG measurements.

The type of texture observed here has a well-documented precedent for small-molecule guests crystallized in cylindrical nanochannels. X-ray diffraction studies of medium-length $n$-alkanes~\cite{Henschel2007}, 1-alcohols~\cite{Henschel2008} and $n$-hexane~\cite{Henschel2009} solidified in mesoporous silicon showed that confined crystallization proceeds by a Bridgman-type mechanism, in which the crystallographic direction of fastest growth becomes aligned with the channel axis, producing a pronounced axial texture while leaving the azimuthal orientation of crystallites in different channels uncorrelated; the general phenomenology of such confinement-controlled crystallization is reviewed in Ref.~\cite{Huber2015}. This is precisely the combination inferred here from the redistribution of $\theta$--$2\theta$ intensities together with the nearly isotropic polarization dependence of the SHG response (Section~\ref{sec:uncorrelated}).

Notably, the same studies found that the degree of axial texture strengthens monotonically with the filling fraction of the channels, approaching a practically perfect texture only for completely filled nanochannels~\cite{Henschel2009}. Since the transmission factors reported in Section~\ref{sec:confinement} ($k_t = 0.016$--$0.025$) indicate substantial void formation, part of the non-monotonic variation of the texture with channel diameter apparent in Figure~\ref{f3} may reflect differences in the degree of filling between membranes rather than confinement as such. 

Overall, the Raman and XRD results confirm that crystalline BZL is formed throughout the investigated range of AAO nanochannel diameters. At the same time, confinement substantially modifies the crystallographic texture of the BZL phase, with the effect depending non-monotonically on the nanochannel diameter. The results obtained for the free-standing membranes are consistent with a crystallization process in which individual nanochannels crystallize independently. Consequently, although the confined BZL crystallites may exhibit preferential orientations with respect to the nanochannel axis, their azimuthal orientations are not mutually correlated across neighboring channels. This distinction between local crystallographic texture and long-range azimuthal correlation is important for understanding the macroscopic nonlinear optical response discussed below.

\subsection{Transmission-Geometry SHG Response of BZL:AAO Nanocomposites}\label{sec:transmission}
\subsubsection{BZL:AAO Nanocomposites with Uncorrelated Crystallization}\label{sec:uncorrelated}

\begin{figure*}[tp]
	\centering
	\includegraphics[height=0.80\textheight]{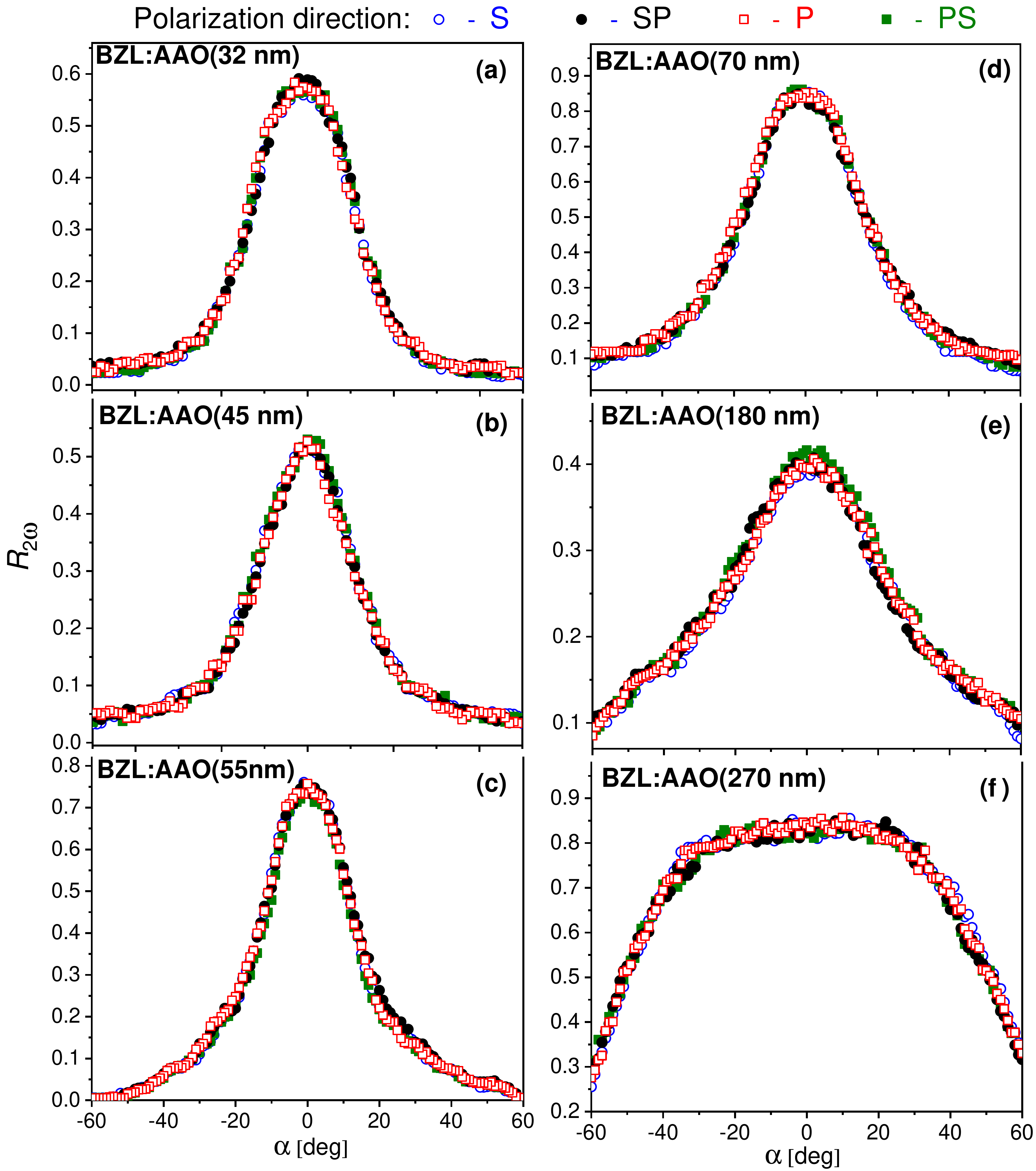}
	\caption{SHG response as a function of the azimuthal angle, $\alpha$, for four fixed incident light polarization directions, S, SP, P, and PS, measured in transmission geometry for alumina--benzil nanocomposite membranes fabricated using the uncorrelated crystallization methodology (free-standing crystallization): BZL:AAO (32~nm) (a), BZL:AAO (45~nm) (b), BZL:AAO (55~nm) (c), BZL:AAO (70~nm) (d), BZL:AAO (180~nm) (e), and BZL:AAO (270~nm) (f).}
	\label{f4}
\end{figure*}

The NLO response of the BZL:AAO nanocomposites was first investigated using free-standing membranes, in which BZL crystallization occurs independently within individual nanochannels. To characterize the SHG response quantitatively, the measured SHG signal from the BZL:AAO nanocomposite membranes was compared with that of a reference $\alpha$-quartz crystal plate measured under identical experimental conditions in the \emph{ooo} coupling geometry, as described in the Experimental Section. The experimental configuration shown in Figure~\ref{f1}(j) is analogous to the Maker-fringe geometry~\cite{Maker1970f} commonly employed in NLO measurements. However, the azimuthal dependences of the SHG response, $R_{2\omega}(\alpha)$, do not exhibit conventional interference fringes (Figure~\ref{f4}), but rather broad angular distributions of diffuse SHG emission. This behavior can be attributed to the strong diffuse multiple scattering in the nanocomposite membranes, which substantially disrupts the phase coherence of the propagating fundamental and SHG fields. Nevertheless, a measurable SHG signal is detected for all investigated samples, confirming that the confined BZL crystalline phase retains its second-order nonlinear optical activity under nanoscale confinement.

The SHG response exhibits a pronounced dependence on the azimuthal angle, with a distinct broad maximum occurring at normal incidence of the pumping beam ($\alpha=0^\circ$). The azimuthal dependences measured for the four incident polarization directions, \textbf{S}, \textbf{SP}, \textbf{P}, and \textbf{PS}, exhibit broadly similar and largely overlapping angular distributions. This behavior indicates that the nonlinear optical response of the individual BZL crystallites is strongly averaged on the macroscopic scale. Although XRD demonstrates that nanoscale confinement can modify the local crystallographic texture of BZL, the absence of a common azimuthal orientation among crystallites formed in neighboring nanochannels prevents this local texture from producing pronounced macroscopic polarization anisotropy.

\begin{figure*}[tp]
	\centering
	\includegraphics[height=0.78\textheight]{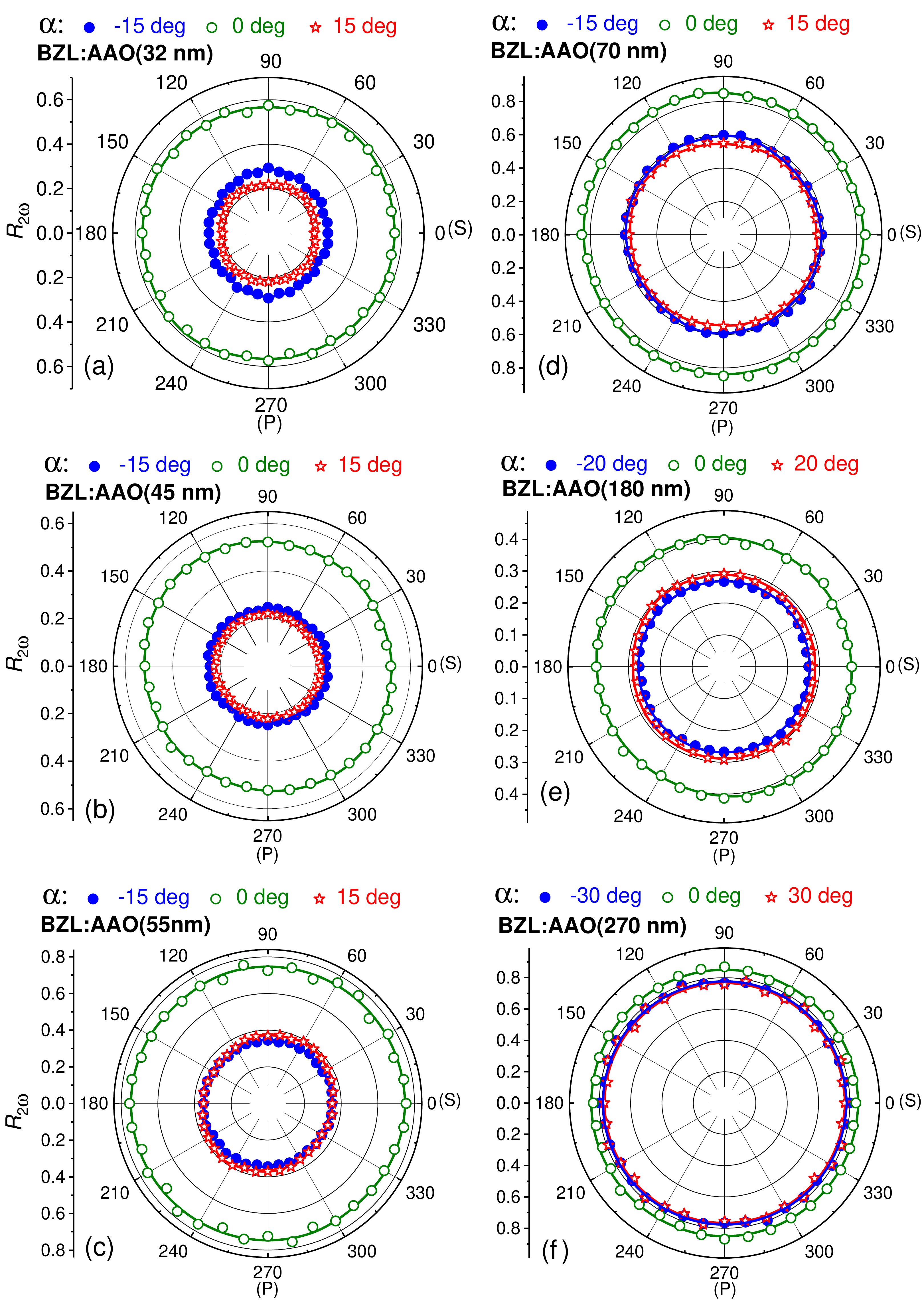}
	\caption{SHG response as a function of the incident-light polarization direction for three fixed azimuthal angles, measured in transmission geometry for alumina--benzil nanocomposite membranes fabricated using the uncorrelated crystallization methodology (free-standing crystallization): BZL:AAO (32~nm) (a), BZL:AAO (45~nm) (b), BZL:AAO (55~nm) (c) and BZL:AAO (70~nm) (d), each at $\alpha=-15^\circ$, $0^\circ$ and $+15^\circ$; BZL:AAO (180~nm) (e) at $\alpha=-20^\circ$, $0^\circ$ and $+20^\circ$; and BZL:AAO (270~nm) (f) at $\alpha=-30^\circ$, $0^\circ$ and $+30^\circ$.}
	\label{f5}
\end{figure*}

An additional feature of the azimuthal SHG response is the systematic variation of its angular width with nanochannel diameter. The relatively narrow distribution observed for the 32 nm membrane (FWHM $\approx$ 26$^\circ$) becomes progressively broader with increasing channel diameter, reaching approximately 110$^\circ$ for the largest channels. This trend can be qualitatively interpreted in terms of aspect-ratio-dependent optical collimation. In nanoporous AAO, the pore length-to-diameter ratio has been shown to substantially influence the optical response, while the columnar pore structure can support waveguiding and introduce pronounced optical anisotropy~\cite{Kumeria2012c,Saito1994o,Noyan2022b}. More generally, direct optical measurements on high-aspect-ratio parallel-hole structures have shown that increasing the ratio of channel height to pore width progressively narrows the angular transmission range, in close agreement with geometrical collimation models~\cite{Westover2022c}. In the present membranes, increasing the channel diameter at comparable membrane thickness reduces the aspect ratio $h/D$, thereby weakening the geometrical constraint on transverse propagation and collection of the generated SHG radiation. Conversely, the higher aspect ratio of the smallest channels provides stronger angular confinement, resulting in a narrower SHG distribution. Although the present data do not allow a quantitative separation of geometrical collimation from diffraction, refraction, and scattering effects, the observed dependence of the FWHM on channel diameter is consistent with an aspect-ratio-dependent collimation of the SHG radiation.

Despite these changes in angular width, the four polarization-dependent responses remain strongly overlapping for each membrane. To further visualize this behavior, the SHG intensities obtained for the \textbf{S}, \textbf{SP}, \textbf{P}, and \textbf{PS} polarization directions are presented in polar coordinates in Figure~\ref{f5}. The resulting patterns are close to circular for all investigated channel diameters, demonstrating that the macroscopic SHG response of the free-standing BZL:AAO membranes is approximately polarization-isotropic. The residual deviations from circular symmetry are relatively small compared with the overall angular distribution and may arise from the local crystallographic texture, the finite distribution of crystallite orientations, and experimental averaging over the illuminated area.

This apparent macroscopic isotropy should not be interpreted as an absence of nonlinear optical anisotropy at the crystallite level. Rather, it results from the spatial averaging of the nonlinear responses of many BZL crystallites with different azimuthal orientations. Thus, the free-standing AAO membranes provide a useful reference state in which confinement modifies the local crystallographic texture while the absence of long-range azimuthal correlation suppresses the corresponding macroscopic nonlinear optical anisotropy. This picture --- a well-defined orientation with respect to the channel axis combined with a random azimuthal orientation from channel to channel --- is exactly the texture expected for Bridgman-type growth in isolated cylindrical nanochannels~\cite{Henschel2007,Henschel2008,Henschel2009,Huber2015}. This distinction provides the basis for comparison with the correlated crystallization regime discussed below.

\subsubsection{Confinement Dependence of the SHG Response}\label{sec:confinement}

The relative SHG intensity, $R_{2\omega}=I_{2\omega}/I_{2\omega}($SiO$_2)$, provides a direct measure of the experimentally detected SHG response but does not by itself represent the intrinsic conversion efficiency of the BZL phase in the composite samples. In particular, capillary voids formed during the crystallization process can introduce considerable multiple light scattering, resulting in a substantial reduction of the SHG signal detected from the BZL:AAO composites. To estimate this effect, the transmission factor $k_t$, which characterizes the attenuation of light associated primarily with scattering, was determined for the individual composite membranes. The factor depends on the composite morphology and membrane thickness $h$. By replacing the picosecond Nd:YAG laser used in the experimental setup (Figure~\ref{f1}(j)) with a low-power continuous-wave laser ($\lambda = 532$~nm), $k_t$ was determined from the ratio of the detected intensities with and without the sample in the optical path. The measured values of $k_t$ were in the range 0.016--0.025. Thus, if the scattering losses associated with the capillary voids could be eliminated, the detected SHG response of uniformly filled BZL:AAO composite membranes could, in this approximation, be enhanced by approximately a factor of 40--60. This observation indicates that more uniform filling of the AAO nanochannels with the organic guest material represents an important route toward improving the effective nonlinear optical performance of the composites. Furthermore, if the embedded organic nanocrystals could be formed with a uniform crystallographic orientation, the resulting nonlinear optical anisotropy could provide an additional means of enhancing or tailoring the SHG response through appropriate selection of the incident geometry and light polarization. A possible route toward such control is correlated capillary crystallization, as demonstrated below.

\begin{figure*}[tp]
	\centering
	\includegraphics[width=0.92\textwidth]{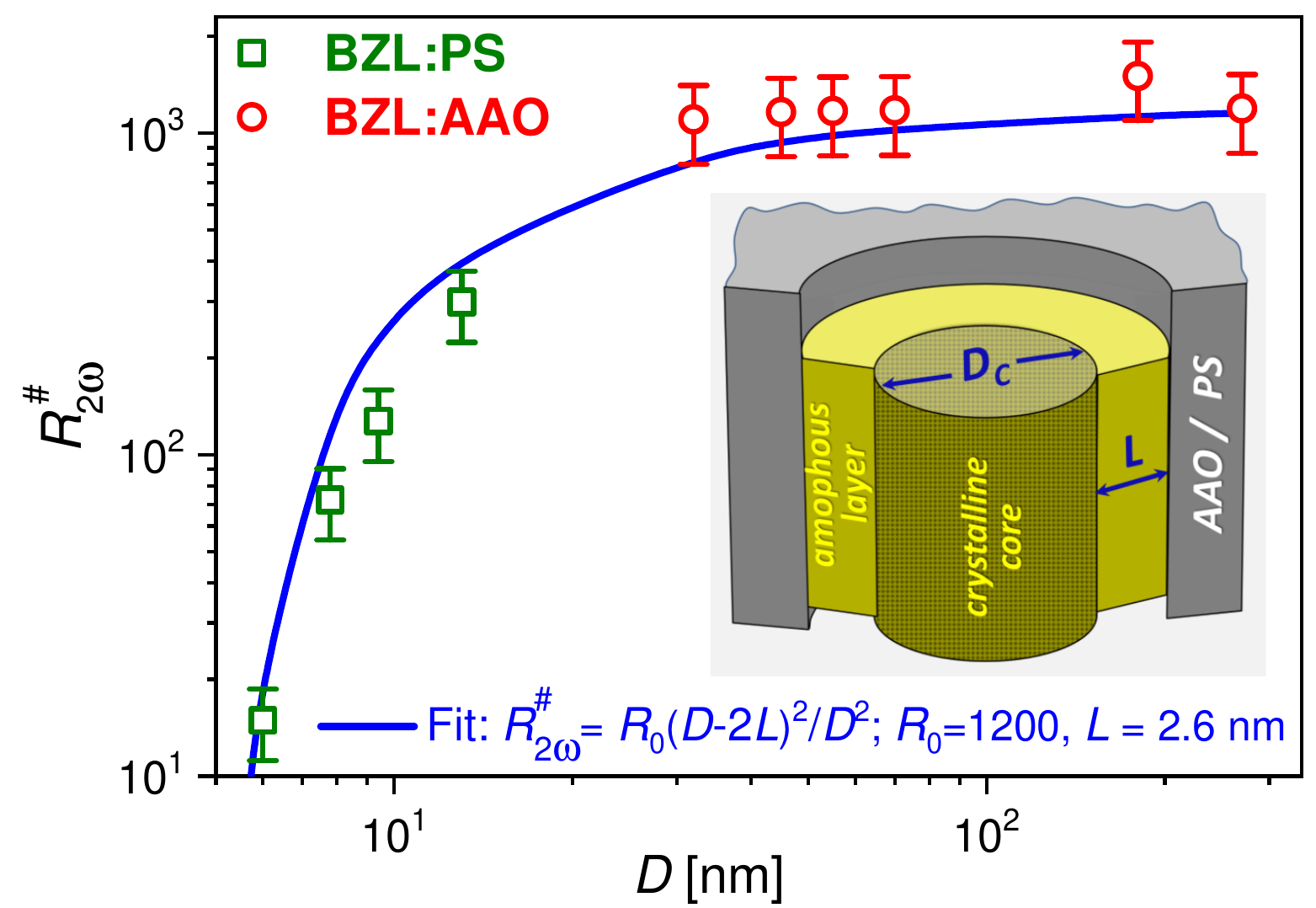}
	\caption{Normalized SHG response $R^{\#}_{2\omega}=(h/1\,\mathrm{mm})^{-1}k_t^{-1}P^{-1}I_{2\omega}/I_{2\omega}($SiO$_2)$ from BZL:PS nanocomposite membranes (squares, green online color)~\cite{Elkarout2023} and BZL:AAO membranes (circles, red online color) as a function of channel diameter $D$. The blue solid line represents the model fit of the $R^{\#}_{2\omega}(D)$ dependence using the function $R_0(D-2L)^2/D^2$, where $R_0=1200$ and $L=2.6$~nm are fitting parameters. The inset schematically illustrates the nanocrystalline cluster morphology consisting of an NLO-inactive amorphous interfacial layer adjacent to the host channel wall and an NLO-active crystalline core.}
	\label{f6}
\end{figure*}

We next consider the direct influence of spatial confinement on the SHG conversion efficiency. In BZL:AAO nanocomposite media, the SHG response originates from the guest BZL crystalline nanoclusters. Following the approach developed in our previous work~\cite{Elkarout2023}, and similarly to the treatment of nonlinear optical response in surface and colloidal systems~\cite{Heinz1983,Jen2009}, the measured SHG intensity of such a multicluster system can be related to the average crystalline cluster volume $V$, the number of nanocrystal clusters $N$ within the path of the incident laser beam, the effective second-order nonlinear polarizability (microscopic hyperpolarizability) of a finite nanocluster $\langle \beta^{(2)}\rangle$, the transmission factor $k_t$ accounting for the reduction of the generated light due to diffuse scattering, and the fundamental light intensity $I_{\omega}$:
\begin{equation}
 I_{2\omega}(\text{BZL:AAO}) \propto k_t\cdot N\cdot V \langle \beta^{(2)}\rangle^2 I_{\omega}^2.
 \label{eq1}
\end{equation}
The number of guest BZL clusters within the path of the pumping laser beam is proportional to the membrane thickness $h$ and the density of nanochannels, represented by the porosity $P$, such that $N \propto P \cdot h$. The volume $V$ in Equation (\ref{eq1}), in the present model, represents the NLO-active part of the BZL cluster and is therefore proportional to the fraction of the crystalline volume within the total cluster volume, $D_c^2/D^2=(D-2L)^2/D^2$. Here, $D_c$ is the diameter of the crystalline core and $L$ is the thickness of the NLO-inactive amorphous BZL layer surrounding the crystalline core, as schematically illustrated in the inset of Figure~\ref{f6}. Accordingly, the relative SHG intensity $R_{2\omega}$, normalized by the membrane thickness $h$, porosity $P$, and transmission factor $k_t$,
\begin{equation}
R^{\#}_{2\omega}=h^{-1}k_t^{-1}P^{-1}R_{2\omega},
\label{eq2}
\end{equation}
provides a convenient quantity for investigating the dependence of the effective SHG response on spatial confinement in the BZL:AAO composite samples. In the simplest geometrical approach, the normalized SHG response $R^{\#}_{2\omega}$ is proportional to the crystalline volume fraction, $D_c^2/D^2$, and can therefore be expressed as
\begin{equation}
 R_{2\omega}^{\#} \propto  \langle \beta^{(2)}\rangle^2\cdot D_c^2/D^2= \langle \beta^{(2)}\rangle^2\cdot (D-2L)^2/D^2= (\beta^{(2)}_{eff})^2
 \label{eq3}
\end{equation}
Within this description, the effective microscopic hyperpolarizability of the finite cluster, reflecting its nonuniform morphology comprising amorphous and crystalline regions, becomes confinement dependent,
\begin{equation}
\beta^{(2)}_{eff}= \langle \beta^{(2)}\rangle(D-2L)/D,
 \label{eq4}
\end{equation}
and consequently gives rise to a confinement-dependent macroscopic SHG response. In the limiting case $D \rightarrow 2L$, the crystalline core vanishes and the model predicts the disappearance of the second-order nonlinear response. In this limit, interfacial disorder prevents the formation of a guest crystalline core in the central part of the channel, leaving the BZL filament effectively amorphous and optically isotropic. Conversely, for $D \gg L$, the contribution of the crystalline core dominates over that of the near-interfacial amorphous layer, and the normalized SHG response approaches its maximum value, with $\beta^{(2)}_{eff} \rightarrow \langle \beta^{(2)}\rangle$. Expressed in the language of the metamaterial picture, the geometrical factor $(D-2L)^2/D^2$ is the filling factor of the individual meta-atom: it is the fraction of each nanochannel cross-section that carries a second-order susceptibility. Since $R^{\#}_{2\omega}$ is already normalized by the membrane thickness and by the porosity, that is, by the number density of meta-atoms, Equation~(\ref{eq3}) states that the squared effective second-order susceptibility of the composite scales with the active volume fraction of its meta-atoms. The interfacial layer $L$ therefore sets a lower bound on the channel diameter for which a nanoporous host can carry a useful effective $\chi^{(2)}$.
To further examine this confinement dependence, we combine the present SHG results for alumina--benzil (BZL:AAO) nanocomposites with the corresponding experimental results obtained in our previous work~\cite{Elkarout2023} for silica--benzil (BZL:PS) nanocomposites. The combined data cover a broad range of channel diameters, from a few nanometers to several hundred nanometers, spanning the nanometric to submicrometric regimes. Remarkably, the simple geometrical model described above provides a reasonable description of the normalized SHG response of both silica--benzil and alumina--benzil nanocomposites despite the simplified assumptions involved. In particular, (i) the magnitude of the second-order nonlinear polarizability $\langle \beta^{(2)}\rangle$ is assumed to have the same average value for all nanocomposite samples, consistent with the essentially disordered crystalline nanocluster texture and the practically isotropic dependence of the SHG response on the pump-light polarization; and (ii) the thickness of the interfacial disordered amorphous BZL layer $L$ is assumed, in the simplest approximation, to be determined by the host--guest interfacial interaction and to remain independent of the channel diameter $D$.

Figure~\ref{f6} presents the normalized SHG response $R^{\#}_{2\omega}$ of the BZL:PS nanocomposite membranes (squares, green online color)~\cite{Elkarout2023} and the BZL:AAO membranes (circles, red online color) as a function of channel diameter $D$. The blue solid line represents the best fit of the $R^{\#}_{2\omega}(D)$ dependence using the function $R_0(D-2L)^2/D^2$, with $R_0$ and $L$ as fitting parameters. The fit yields an effective thickness of the amorphous interfacial layer of $L=2.6$~nm. Within the accuracy of the present analysis, this value is consistent with the corresponding value of $L=2.5$~nm obtained for silica--benzil nanocomposites in Ref.~\cite{Elkarout2023}. The present data therefore reproduce, in a chemically different host and over a channel-diameter range more than an order of magnitude wider, the interfacial layer thickness established for mesoporous silica. We regard this agreement as the principal significance of the confinement analysis: an octyl-grafted, hydrophobic silica wall and a bare, hydroxylated alumina wall present very different interfacial chemistries, so the coincidence of $L$ argues that the thickness of the non-crystalline interfacial region is governed by the packing frustration of benzil at a rigid wall rather than by specific guest--host interactions. Within the meta-atom picture, $L$ thus acts as a transferable material parameter that sets the smallest channel diameter from which a molecular nonlinear optical metamaterial can be built.

It should be emphasized that Equation (\ref{eq3}) represents a simplified geometrical description. It assumes a diameter-independent effective interfacial layer thickness and does not explicitly account for possible changes in crystallite size, crystallographic texture, molecular orientation, or phase composition with confinement. Nevertheless, the ability of this simple model to reproduce the observed dependence over a broad range of channel diameters suggests that the increasing relative contribution of an interfacial BZL region provides a physically meaningful first-order description of the confinement-induced suppression of the second-order NLO response.

\subsubsection{BZL:AAO Nanocomposites with Correlated Crystallization}\label{sec:correlated}

\begin{figure*}[tp]
	\centering
	\includegraphics[width=0.82\textwidth]{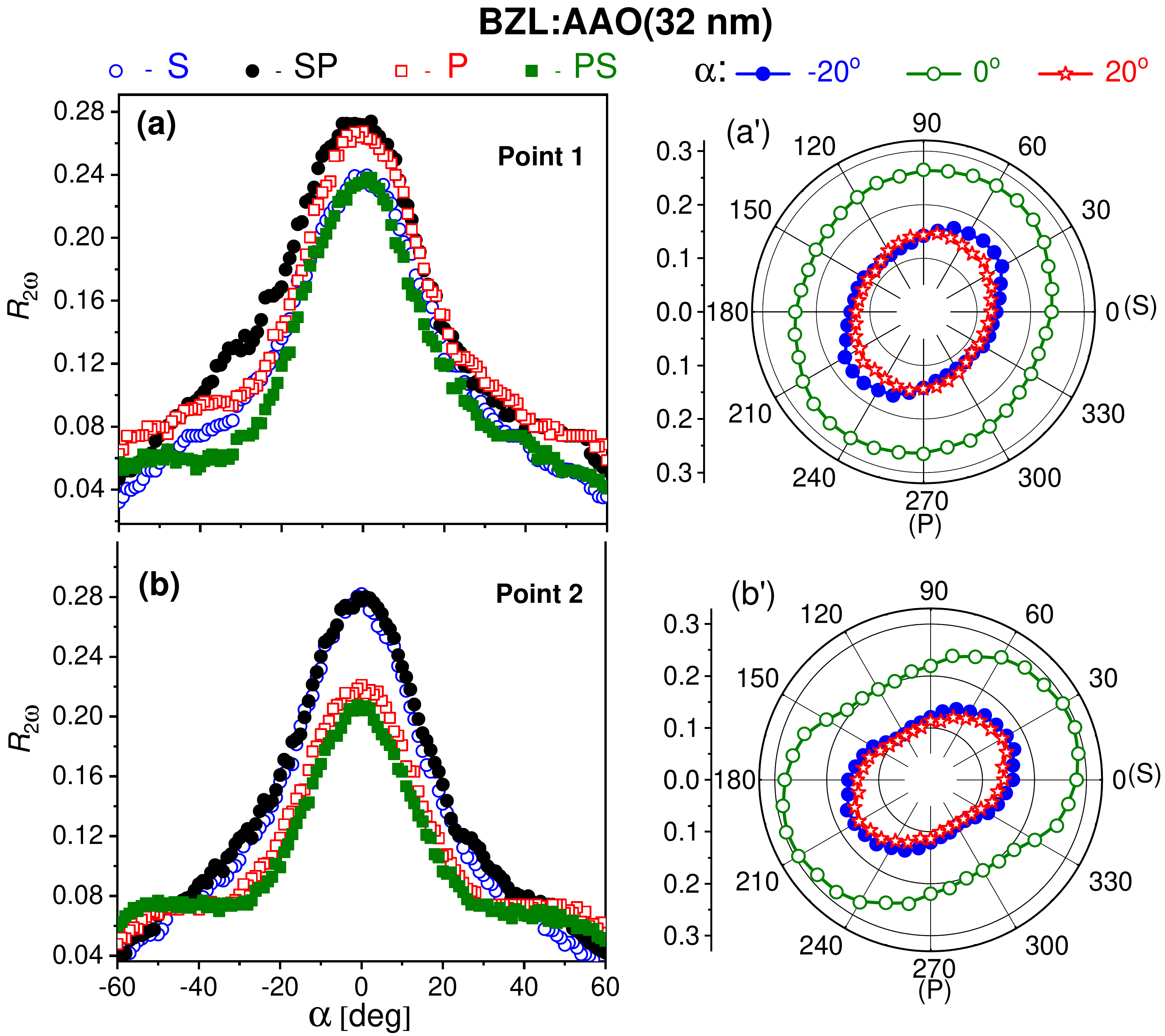}
	\caption{SHG response as a function of the azimuthal angle, $\alpha$, measured at two different local sample positions relative to the focused incident laser beam (panels (a) and (b)) for four fixed incident-light polarization directions, S, SP, P, and PS, in an alumina–benzil nanocomposite membrane, BZL:AAO(32 nm), fabricated using the correlated crystallization methodology. Panels (a') and (b') show the corresponding SHG response as a function of the incident-light polarization direction for three fixed azimuthal angles, $\alpha=-20^\circ$, $0^\circ$, and $20^\circ$, measured at the same two local positions of the incident beam.}
	\label{f7}
\end{figure*}

\begin{figure*}[tp]
	\centering
	\includegraphics[width=0.82\textwidth]{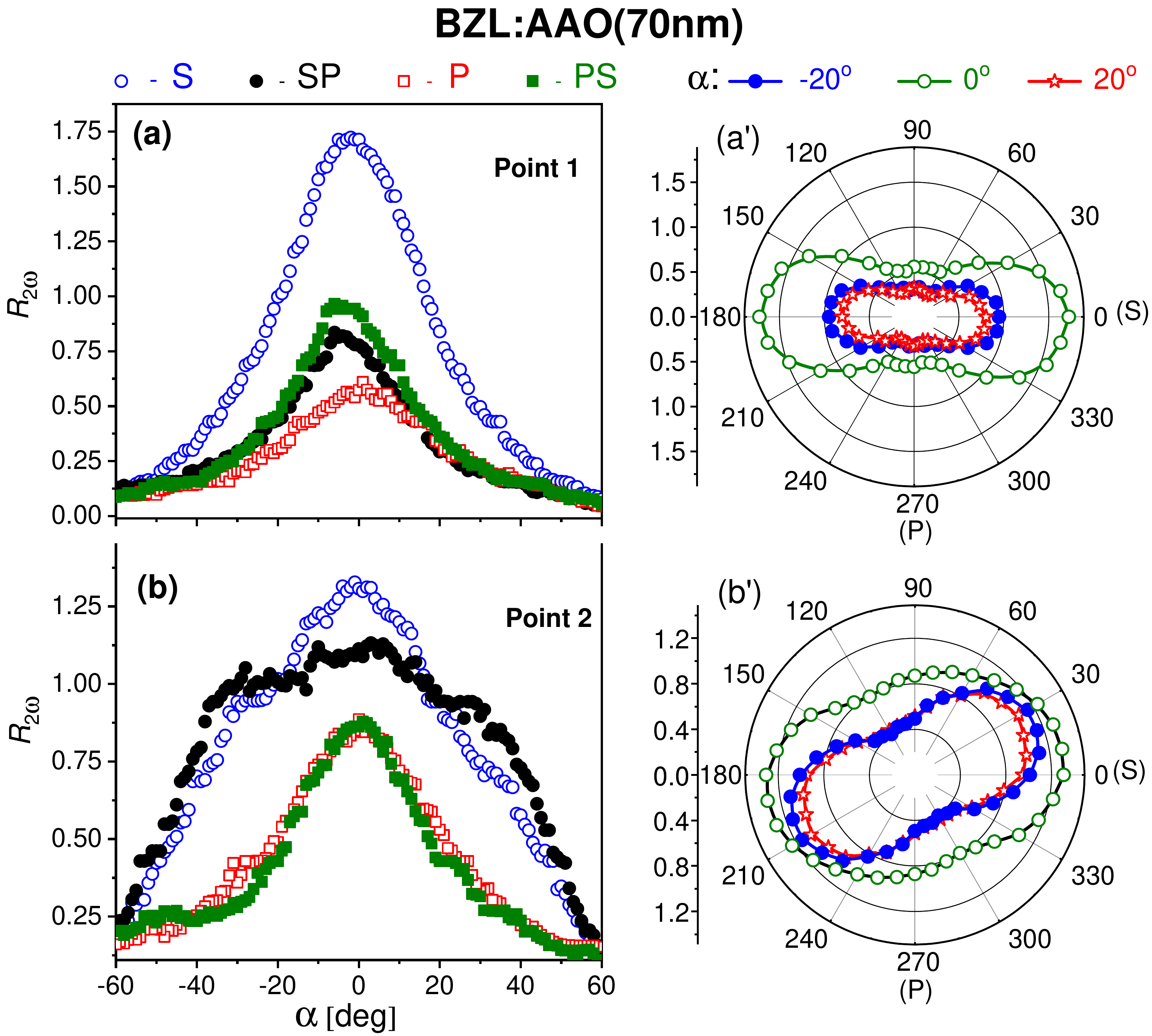}
	\caption{SHG response as a function of the azimuthal angle, $\alpha$, measured at two different local sample positions relative to the focused incident laser beam (panels (a) and (b)) for four fixed incident-light polarization directions, S, SP, P, and PS, in an alumina–benzil nanocomposite membrane, BZL:AAO(70 nm), fabricated using the correlated crystallization methodology. Panels (a') and (b') show the corresponding SHG response as a function of the incident-light polarization direction for three fixed azimuthal angles, $\alpha=-20^\circ$, $0^\circ$, and $20^\circ$, measured at the same two local positions of the incident beam.}
	\label{f8}
\end{figure*}

We now turn to BZL:AAO nanocomposites prepared by correlated capillary crystallization, in which solidification is initiated in the bulk BZL layer between the membrane and the glass substrate and subsequently propagates into the nanochannels (Figure~\ref{f1}(g)). The effect is illustrated for two representative samples, BZL:AAO(32~nm) (Figure~\ref{f7}) and BZL:AAO(70~nm) (Figure~\ref{f8}). For each sample, panels (a) and (b) show the SHG response as a function of the azimuthal angle $\alpha$ for the four incident polarization directions S, SP, P and PS, recorded at two different positions of the focused pump beam (spot diameter $d = 0.3$~mm) on the same membrane; panels (a$'$) and (b$'$) show the corresponding polarization dependences at three fixed azimuthal angles, $\alpha = -20^{\circ}$, $0^{\circ}$ and $+20^{\circ}$.
The contrast with the free-standing membranes is immediate. Whereas uncorrelated crystallization yields polarization dependences of nearly circular shape (Figure~\ref{f5}), the correlated composites exhibit distinctly elongated, strongly anisotropic patterns in both the azimuthal and the polarization dependences. The confined BZL crystallites therefore retain a common crystallographic orientation over length scales large compared with the interchannel distance, and their microscopic nonlinear responses no longer average to an isotropic macroscopic response.

The anisotropy is, however, not uniform across a given sample: both its magnitude and the direction of its principal axis change markedly between the two measurement positions in Figures~\ref{f7} and~\ref{f8}. This is the expected consequence of the templating mechanism. Solidification of the interfacial bulk layer produces a polycrystalline texture whose domains extend over submillimeter-to-millimeter distances; each domain then acts as a common crystallographic template for the very large number (of order $10^{7}$) of nanochannels beneath it, so that the composite consists of a mosaic of macroscopically correlated regions with mutually different orientations. Because the pump spot ($d = 0.3$~mm) is comparable to or smaller than the domain size, each measurement position probes essentially a single template domain, and the observed principal axes follow the local domain orientation rather than any sample-fixed direction. A spatially resolved areal scan would map this mosaic directly and is a natural next step.

It should be stressed that this mosaic structure also limits what diffraction can contribute here. The X-ray measurements of Section~\ref{sec:texture} were performed with a laboratory diffractometer whose 1~cm long gap diaphragm illuminates an area exceeding the size of a single correlated domain by more than an order of magnitude. A $\theta$--$2\theta$ scan of a correlated sample therefore averages over many neighboring domains of different azimuthal orientation and would, by construction, return a texture close to that of the free-standing membranes, irrespective of how well ordered each individual domain is. The diffraction and SHG experiments can moreover not be performed on the same local spot of the sample. With its focused probe ($d = 0.3$~mm), SHG is in this respect the more local of the two techniques, and the point-to-point variation of the SHG anisotropy documented in Figures~\ref{f7} and~\ref{f8} is the direct experimental signature of a composite that is ordered on the submillimeter scale but disordered on the centimeter scale. A structural verification resolved at the domain level would require a microfocused X-ray beam --- scanning micro- or nanodiffraction at a synchrotron source --- which we regard as the natural next step.

The mechanism underlying the correlated crystallization is not specific to benzil. Steinhart et al. reported that polymers crystallized in AAO templates form macroscopic arrays with uniform crystal orientation precisely when a bulk-like surface film connects the individual nanostructures, the dominant growth direction being aligned with the long axes of the template pores~\cite{Steinhart2006}. The interfacial BZL layer employed here plays the same role, and the free-standing and substrate-supported membranes of the present study correspond to the absence and the presence of such a connecting layer. What is new here is the consequence for the second-order optical response: because SHG is sensitive to the orientation of every contributing crystallite, the presence or absence of a connecting bulk layer switches the macroscopic nonlinear response between an essentially isotropic and a strongly anisotropic regime. The mosaic structure of the seed layer, in turn, sets the length scale over which this control extends and thus the practical limit on the uniformity that can be achieved.

Preparing membranes that are correlated uniformly over the entire sample area remains technologically demanding. The most promising route is to replace the polycrystalline interfacial layer by a single-crystalline BZL seed layer, so that one crystallographic orientation is imprinted across the whole membrane. This lies outside the scope of the present work, but it is worth stating what it would enable a nanocomposite that combines a well-defined macroscopic second-order susceptibility tensor with the form birefringence of the nanochannel array. Since the effective refractive indices of such composites can be tuned through porosity and filling fraction~\cite{Kumeria2012c,Noyan2022b}, this opens a route towards birefringent phase matching in a material whose nonlinear response is provided by an organic guest --- a combination that is difficult to realize in bulk molecular crystals and that is a prerequisite for efficient frequency conversion.

\subsection{Subwavelength Limit of the Meta-Atom Description}\label{sec:subwavelength}

The metamaterial description outlined above is quantitative only as long as the lattice constant of the nanochannel array is small compared with the wavelength of the light inside the composite. For the hexagonal AAO lattice the interpore distance $D_{\mathrm{int}}$ follows from the channel diameter $D$ and the porosity $P$ of Table~\ref{tab1}, and ranges from approximately 65 nm for AAO(32) to approximately 120 nm for AAO(70), but reaches several hundred nanometers for the two widest membranes. Referred to the second-harmonic wavelength inside the composite, $\lambda_{2\omega}/n \approx 330$ nm for $n \approx 1.6$, the four narrow membranes correspond to $D_{\mathrm{int}}/(\lambda_{2\omega}/n) \approx 0.2$--$0.4$ and are thus firmly subwavelength, whereas AAO(180) and AAO(270) reach or exceed unity and therefore fall outside the effective-medium regime.

The sample series investigated here consequently spans the boundary of the metamaterial description rather than lying entirely on one side of it. As shown below, several of the optical observations change character precisely at this boundary, which supports the effective-medium interpretation for the narrow channels and marks its limit for the wide ones. We note that the interpore distances quoted above are derived from $D$ and $P$ assuming an ideal hexagonal arrangement; they are consistent with the electron micrographs of Figure~\ref{f1}(a--f).

\subsection{Reflection-Geometry SHG Response of BZL:AAO Nanocomposites}\label{sec:reflection}

An alternative approach to characterizing the nonlinear optical (NLO) response of crystalline and composite materials is provided by SHG measurements in a reflection geometry (Figure~\ref{f1}(l)). In this configuration, a layer of compressed microcrystalline $\alpha$-SiO$_2$ powder, approximately $150~\mu$m thick and positioned at an azimuthal angle of $\alpha=45^\circ$, was used as the reference. The photomultiplier (PM) was fixed at $90^\circ$ with respect to the incident laser beam, while the sample was rotated through $360^\circ$ in $1^\circ$ increments, starting from its normal position ($\alpha=0^\circ$).

\begin{figure*}[tp]
	\centering
	\includegraphics[width=0.86\textwidth]{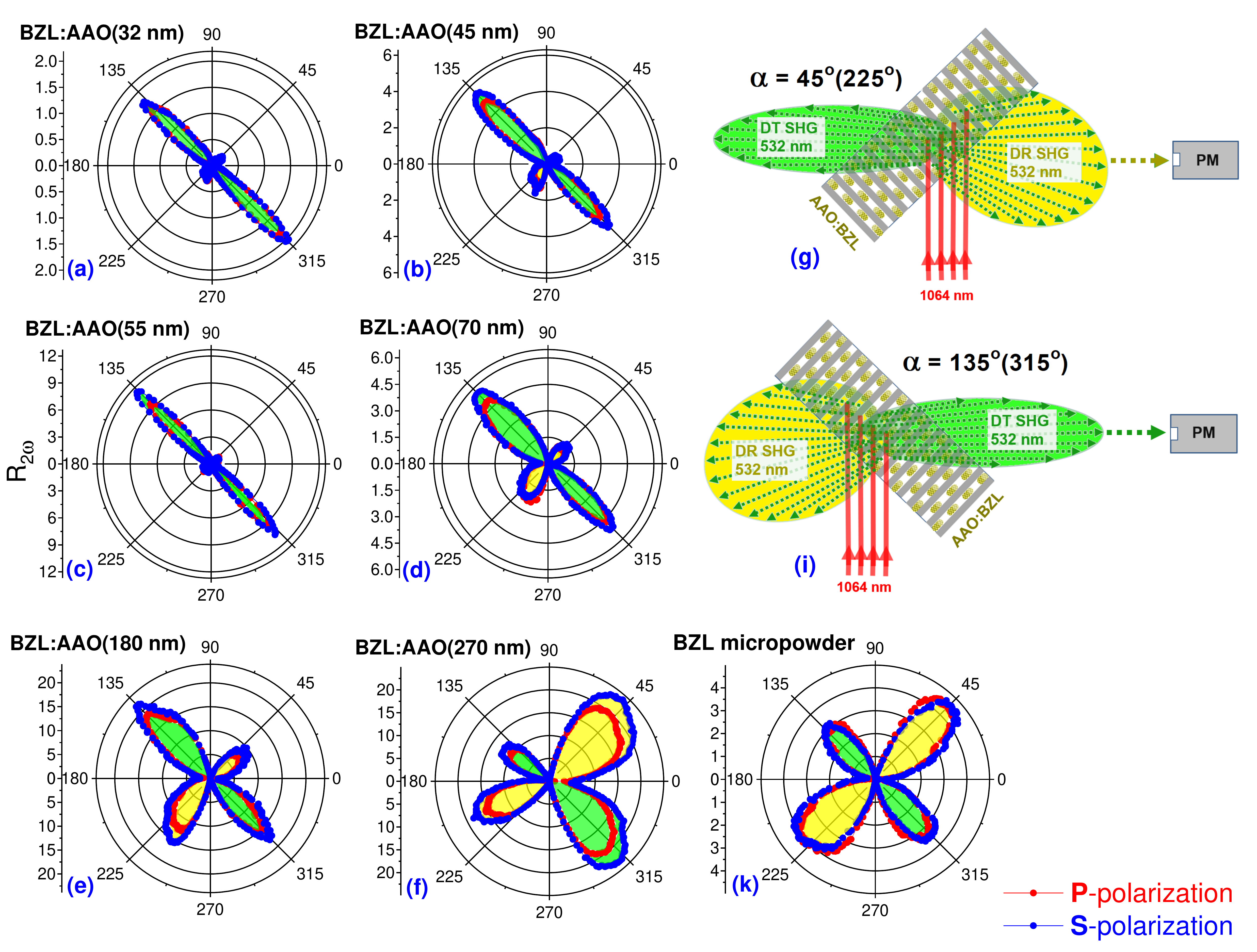}
	\caption{SHG response as a function of the azimuthal angle, $\alpha$, measured in the 90$^\circ$-reflection geometry for four fixed incident-light polarization directions, S, SP, P, and PS, for alumina–benzil nanocomposite membranes fabricated using the uncorrelated crystallization methodology (free-standing crystallization): BZL:AAO (32 nm) (a), BZL:AAO (45 nm) (b), BZL:AAO (55 nm) (c), BZL:AAO (70 nm) (d), BZL:AAO (180 nm) (e), and BZL:AAO (270 nm) (f). The azimuthal SHG response of a 150$\mu$-thick BZL micropowder sample (k) is shown for comparison. Green and yellow regions indicate the azimuthal ranges corresponding to diffuse-transmitted (DT) and diffuse-reflected (DR) SHG emission, respectively. Panels (g) and (i) illustrate two detection regimes for DT and DR SHG emission, respectively, realized at two different azimuthal sample positions relative to the incident laser beam and the PM detector, as indicated in the schematics.}
	\label{f9}
\end{figure*}

Importantly, different angular sectors correspond to distinct regimes of the detected SHG emission. For sample positions within the angular ranges $0^\circ$--$90^\circ$ and $180^\circ$--$270^\circ$, the detected signal is associated with diffuse reflected (DR) SHG emission, whereas positions within $90^\circ$--$180^\circ$ and $270^\circ$--$360^\circ$ correspond to diffuse transmitted (DT) SHG emission. These two detection regimes are schematically illustrated in the insets of Figure~\ref{f9}(g) and (i), respectively. Figures~\ref{f9}(a)--(f) present polar plots of the relative SHG response, $R_{2\omega}$, as a function of the incident angle $\alpha$ for BZL:AAO nanocomposites under \textbf{P}- and \textbf{S}-polarized excitation. The corresponding azimuthal dependence measured for a compressed microcrystalline BZL layer is shown in Figure~\ref{f9}(k) for comparison.

In conventional, strongly scattering microcrystalline media, multiple scattering limits the penetration depth of both the fundamental and generated SHG radiation. Consequently, the DT SHG signal is generally expected to be substantially weaker than the DR signal. This behavior is observed for the compressed microcrystalline BZL layer, for which the DR and DT contributions are indicated in yellow and green, respectively, in Figure~\ref{f9}(k). In marked contrast, the BZL:AAO membranes exhibit an unusual response in which the DT SHG emission dominates over the DR contribution (green and yellow, respectively, in Figures \ref{f9}(a)--(f)). This reversal of the relative emission intensities distinguishes the nanocomposite membranes from conventional opaque, diffusely scattering media.

The relative strength of the DT and DR signals also depends on the nanochannel diameter. For smaller diameters ($D\leq70$ nm), the DT-to-DR intensity ratio increases as the channel diameter decreases. In this range, the DT emission exhibits a relatively narrow azimuthal distribution, with an angular width of approximately $\Delta\alpha\sim13^\circ$--$20^\circ$ and a maximum near $\alpha\approx135^\circ$ and $315^\circ$. The maximum DT response exceeds the corresponding DR maximum by a factor that increases from approximately 3 to 10 as the channel diameter decreases. For larger channel diameters, the DT and DR signals become increasingly comparable, as is particularly evident for $D=180$ nm and, even more clearly, for $D=270$ nm. Moreover, a pronounced asymmetry emerges between measurements performed at inverse angular positions, $\alpha$ and $\alpha+180^\circ$, showing that the detected SHG response depends on which face of the membrane is turned towards the incident beam. Such a front--back asymmetry cannot arise from an ideal array of straight cylindrical channels, but it follows naturally from the fact that AAO nanochannels are slightly conical. Unidirectional electrochemical anodization produces a systematic taper along the growth direction, which has been quantified for membranes of the present type by nondestructive imbibition interferometry, cross-sectional SEM and FIB tomography: for a nominally 65~nm, 108~$\mu$m thick membrane the three methods consistently yield a conicity parameter $H = r_{f}/r_{0} = 0.82$--$0.88$, i.e.\ a reduction of the channel radius by 12--18\% between the mouth and the base~\cite{Cencha2019}. In the present membranes the taper is expected to be at least as large, since the porosities listed in Table~\ref{tab1} (15--58\%) lie well above the $\approx 10$\% characteristic of as-anodized self-ordered alumina~\cite{Nielsch2002} and therefore imply a post-anodization pore-widening etch, which acts longest at the channel mouth and thus enhances the conicity; the deliberate tailoring of AAO nanopore geometry has recently been explored further~\cite{Liu2025}. A second, related source of front--back asymmetry is the branched, dendritic pore structure left near the former barrier layer when the membrane is opened chemically.

Channel conicity breaks the symmetry of the composite in two ways that are directly relevant here. First, a tapered channel is a tapered waveguide: its angular acceptance, and the divergence of the radiation it delivers, differ according to whether light enters at the wide or at the narrow end, so that both the DT/DR ratio and the angular width of the emission lobe are expected to depend on the membrane orientation. Second, the local degree of confinement --- and hence the local ratio of NLO-active crystalline core to NLO-inactive interfacial shell --- varies along the channel axis, so that the SHG source distribution is itself non-uniform in depth. The latter also provides a natural explanation for the differences in relative XRD intensities recorded from opposite sides of the same membrane (Section~\ref{sec:texture}), since in the $\theta$--$2\theta$ geometry the two measurements weight the regions near the wide and near the narrow end of the channels differently. We note, however, that for the extraction of the interfacial layer thickness $L$ in Section~\ref{sec:confinement} the effect on $L$ is expected to be small compared with the scatter of the data in Figure~\ref{f6}.

The origin of this unusual SHG behavior remains to be established. In a recent study of mesoporous SiO$_2$--LiNbO$_3$ nanocomposite membranes~\cite{Shchur2024s}, a pronounced, angularly confined DT SHG response was likewise observed in the $90^\circ$ reflection geometry. In that work, the anomalous emission was tentatively attributed to internal reflections within the tubular nanochannel network. The similarity between the SHG responses observed for the SiO$_2$--LiNbO$_3$~\cite{Shchur2024s} and the present BZL membranes suggests that the porous-host architecture may play an important role in determining the angular distribution of the detected SHG signal.

In particular, the pronounced and angularly confined DT SHG emission observed in BZL:AAO membranes with smaller channel diameters ($D\leq50$ nm) occurs near $\alpha\approx135^\circ$ and $315^\circ$, corresponding to an incidence angle of approximately $45^\circ$ with respect to the tubular channel network. This angular correspondence is consistent with the possibility that internal side-wall reflections and morphology-dependent light propagation within the nanochannels contribute to the enhanced DT SHG response. The dependence on channel diameter, together with the asymmetry between inverse sample orientations, may also reflect the influence of channel geometry and the spatial distribution of the confined BZL crystallites. These interpretations remain tentative, however, and further theoretical and experimental investigations are needed to establish the mechanisms responsible for the observed behavior. It is, however, worth noting where in the sample series this behavior occurs. The pronounced dominance of diffuse-transmitted over diffuse-reflected emission, and the narrow angular width of the transmitted lobe, are found for $D \leq 70$ nm, that is, exactly for those membranes whose interpore distance is well below the second-harmonic wavelength inside the composite (Section~\ref{sec:subwavelength}). For AAO(180) and AAO(270), which fall outside the effective-medium regime, the two contributions become comparable and the angular distribution broadens by a factor of four. The anomalous directional emission is therefore confined to the metamaterial limit of the series, which suggests that it originates in the collective optical response of the subwavelength channel array rather than in the properties of the individual BZL crystallites.

Despite the need for a more detailed explanation, the present results establish distinctive angular and directional features of the SHG response in BZL:AAO membranes. In particular, the measurements extend the investigation across channel diameters spanning the nanoscale to submicron range, providing a broader basis for comparison with the previously reported silica--lithium niobate nanocomposites~\cite{Shchur2024s}. Beyond their fundamental interest, these findings also motivate further investigation of the directional SHG emission from such nanocomposites in the context of potential nonlinear-optical applications.

\section{Conclusion}

In this work, benzil--anodic aluminum oxide (BZL) nanocomposites were investigated as a model system for exploring the interplay between spatial confinement, crystallization pathway, and nonlinear optical response in molecular nanocomposites. AAO membranes with nanochannel diameters ranging from 32 to 270~nm were employed, providing a substantially broader confinement range than that available in our previous studies of BZL confined in mesoporous silica. Raman spectroscopy and X-ray diffraction confirmed that crystalline BZL is formed throughout the investigated range of channel dimensions, while the redistribution of diffraction intensities demonstrates that confinement modifies the crystallographic texture of the embedded molecular phase.

For free-standing membranes, in which crystallization proceeds independently within individual nanochannels, the SHG response is characterized by broad angular distributions and an approximately polarization-isotropic macroscopic response. This behavior indicates that the nonlinear responses of individual BZL crystallites are largely averaged because their azimuthal crystallographic orientations are not correlated over macroscopic distances. At the same time, the angular width of the diffuse SHG emission systematically increases with increasing channel diameter, consistent with a reduction of geometrical angular confinement as the membrane aspect ratio decreases.

The magnitude of the SHG response exhibits a pronounced dependence on channel diameter. By combining the present BZL results with our previous data for BZL confined in porous silica, a simple geometrical model based on an NLO-active crystalline core surrounded by an NLO-inactive interfacial layer provides a consistent description over a broad range of channel dimensions. The resulting effective interfacial-layer thickness, $L\approx2.6$~nm, is close to the value previously obtained for BZL--silica nanocomposites. This agreement suggests that the relative volume of the crystalline core, rather than the host material alone, provides an important first-order parameter governing the confinement-induced variation of the effective SHG response. It should be stated explicitly that this part of the present study confirms and extends, rather than establishes, the behavior reported for benzil confined in mesoporous silica~\cite{Elkarout2023}; its value lies in demonstrating that the core--shell description and the interfacial layer thickness carry over to a chemically distinct host and to channel diameters an order of magnitude larger. The genuinely new element of this work is the control of the macroscopic orientation of the confined crystallites through the crystallization pathway, and the resulting switching of the effective second-order response between an isotropic and a strongly anisotropic regime. The analysis also indicates that diffuse scattering associated with incomplete filling and crystallization-induced voids substantially reduces the experimentally detected SHG intensity.

A distinctly different behavior is obtained when crystallization is initiated by a bulk BZL layer formed at the AAO--substrate interface. In this correlated crystallization regime, the crystallization front propagates from the interfacial BZL layer into the nanochannels, producing spatially correlated BZL crystallites over submillimeter-to-millimeter length scales. The resulting nanocomposites exhibit pronounced azimuthal and polarization-dependent SHG anisotropy, in contrast to the nearly isotropic response of the free-standing membranes. The strong variation of the anisotropic response from one position to another is consistent with the polycrystalline texture of the interfacial BZL layer, whose differently oriented domains act as local crystallographic templates for the confined phase. These results demonstrate that the crystallization pathway provides an independent means of controlling the macroscopic nonlinear optical anisotropy, in addition to the effect of spatial confinement on the magnitude of the response. The composites are thus locally ordered on the submillimeter scale and disordered on the centimeter scale, a combination that the focused SHG probe resolves but that a laboratory diffraction experiment averages out.

The $90^\circ$-reflection measurements reveal a further distinctive feature of the BZL system. For the smaller nanochannels, particularly for $D\leq70$~nm, the diffuse-transmitted SHG emission can substantially exceed the diffuse-reflected contribution and is concentrated within relatively narrow angular regions near $\alpha\approx135^\circ$ and $315^\circ$. The characteristic emission direction, its dependence on channel diameter, and the observed asymmetry between inverse membrane orientations point to an important role of the porous-channel architecture in determining the propagation and angular distribution of the generated SHG radiation. Internal reflections within the nanochannels represent one possible contribution, consistent with related observations in porous SiO$_2$--LiNbO$_3$ nanocomposites, although the mechanism cannot yet be established conclusively.

Overall, the results demonstrate two complementary structural routes for engineering the nonlinear optical response of molecular nanocomposites: nanoconfinement controls the effective magnitude of the SHG response through the relative contribution of the crystalline phase, whereas correlated crystallization provides a means of introducing macroscopic nonlinear optical anisotropy through spatial organization of the crystallites. The combination of these effects, together with the unusual directional diffuse SHG emission observed in AAO, identifies porous BZL membranes as a versatile platform for studying and potentially engineering nonlinear optical functionality in confined molecular materials. Further progress toward practical applications will require improved control of channel filling and crystallographic templating, as well as a detailed theoretical description of light propagation, scattering, and SHG generation within the nanochannel network. Read in the language of photonic metamaterials, the nanochannel array constitutes a subwavelength lattice of meta-atoms, the interfacial layer fixes their active volume fraction, and the crystallization pathway programs the orientation of their nonlinear response. Confined crystallization thus provides a processing route to nonlinear optical metamaterials whose effective susceptibility tensor is written during fabrication rather than imposed by the choice of material.

Beyond their demonstrated optical-frequency SHG response, BZL nanocomposites may offer a platform for exploring structure–property relationships in the terahertz (THz) frequency range. The combination of a porous aluminum oxide framework, nanoconfined benzil crystallites, and interfacial regions provides a means of investigating how pore geometry and crystallization pathway influence the effective dielectric response, polarization anisotropy, and THz absorption of organic–inorganic nanocomposites. Such characteristics could be relevant to polarization-sensitive THz components, dielectric sensing, and hybrid THz devices. Furthermore, the second-order nonlinear optical activity of crystalline benzil motivates future investigations of possible optical-rectification and THz-generation processes in appropriately designed composites. These applications remain prospective and require direct THz spectroscopic and nonlinear measurements to establish the relevant dielectric and frequency-dependent nonlinear properties.

The architecture also suggests a route to a response that can be addressed after fabrication. Crystallization in the nanochannels is not space-filling: residual capillary voids remain between the BZL crystallites and the hydroxylated alumina walls, and it is the refractive-index contrast at these voids that underlies the diffuse SHG emission reported above. Because BZL is hydrophobic, insoluble in water and forms no hydrates, water can be condensed into and desorbed from this void space reversibly without perturbing the crystalline guest. Filling the voids with water ($n \approx 1.33$) reduces the index contrast against the alumina host and should therefore shift the effective permittivity tensor and the form birefringence of the composite, modify the local-field factors entering the effective $\chi^{(2)}$, and redistribute the second-harmonic emission from diffuse towards coherent. Since the capillary condensation pressure scales inversely with pore size, the present series of channel diameters offers a built-in range of switching humidities. A nonlinear optical metamaterial addressed by water would connect the confinement- and pathway-based control demonstrated here to the emerging design of water-driven materials for the control of electromagnetic waves.

\section{Experimental Section}

\subsection{Preparation and Characterization of BZL:AAO Nanocomposites}

Nanoporous alumina membranes, \emph{p}Al$_2$O$_3$ (hereafter denoted as \emph{AAO}), were purchased from Smart Membranes GmbH (Halle, Germany). The nominal specifications of the \emph{AAO} membranes were verified by scanning electron microscopy (SEM; Figure~\ref{f1}(a--f)), yielding average nanochannel diameters ranging from 32 to 270~nm and porosities between 15 and 58\%, as summarized in Table~\ref{tab1}. The SEM images reveal pronounced differences in the lateral ordering of the nanochannels. The membrane shown in Figure~\ref{f1}(f) exhibits a relatively irregular pore arrangement, characteristic of AAO templates prepared by one-step anodization, whereas the membranes shown in Figure~\ref{f1}(a--e) exhibit a highly ordered, approximately hexagonal pore arrangement characteristic of two-step anodization. Despite these differences in lateral ordering, all investigated AAO membranes consist of approximately parallel cylindrical nanochannels oriented perpendicular to the membrane surface. This morphology defines a well-defined macroscopic structural axis and provides a suitable geometrical framework for investigating anisotropic optical responses of the resulting nanocomposites.

\begin{table}[tb]
\caption{\label{tab1}Structural parameters of the AAO membranes used for the fabrication of BZL:AAO nanocomposites, including the average nanochannel diameter ($D$), porosity ($P$), and membrane thickness ($h$).}
\begin{ruledtabular}
\begin{tabular}{cccc}
AAO & $D$ [nm] & $P$ [\%] & $h$ [$\mu$m] \\
\colrule
AAO(32) & 32$\pm$2 & 22$\pm$2.0 & 100$\pm$5 \\
AAO(45) & 45$\pm$3 & 17$\pm$1.5 & 120$\pm$5 \\
AAO(55) & 55$\pm$4 & 26$\pm$2.5 & 100$\pm$5 \\
AAO(70) & 70$\pm$5 & 30$\pm$3.0 & 140$\pm$5 \\
AAO(180) & 180$\pm$10 & 15$\pm$1.5 & 110$\pm$5 \\
AAO(270) & 270$\pm$20 & 58$\pm$4.0 & 65$\pm$5 \\
\end{tabular}
\end{ruledtabular}
\end{table}

Prior to infiltration, the AAO membranes were annealed at 440~K for approximately 15~min to remove adsorbed water from the channel walls. The membranes were then filled by capillary action (spontaneous imbibition) with molten benzil (BZL) in its isotropic liquid phase at a temperature slightly above the bulk crystallization temperature ($T_m \approx 369$~K). Subsequent slow cooling at a rate of approximately 1~K$\cdot$min$^{-1}$ induced crystallization of the confined BZL. As previously observed for BZL confined in porous silica matrices, crystallization under nanoconfinement can produce a heterogeneous morphology comprising a disordered interfacial region adjacent to the channel walls and a more ordered crystalline core~\cite{Elkarout2023}. The resulting composites are hereafter denoted BZL:AAO and consist of an AAO host containing confined BZL crystalline phases that provide the second-order optical response associated with the noncentrosymmetric crystal structure of benzil.

Two distinct preparation approaches were employed to realize different crystallization regimes, referred to as correlated and uncorrelated crystallization. In the free-standing AAO membranes (Figure~\ref{f1}(h)), solidification occurs independently within individual nanochannels. Consequently, the crystallographic orientations of BZL nanocrystals formed in neighboring channels are not mutually constrained and can vary independently with respect to both the nanochannel axis and the lateral reference directions. This results in an ensemble of nanocrystals with essentially uncorrelated crystallographic orientations.

In the second approach, the AAO membrane was placed on a glass substrate with a narrow gap between the membrane and substrate, which was filled with molten BZL (Figure~\ref{f1}(g)). This configuration promotes a correlated crystallization process. Because the crystallization temperature of BZL in the bulk interfacial layer is substantially higher than that under nanoconfinement, crystallization initially takes place in the intermediate bulk layer. Upon further cooling, crystallization proceeds from this layer into the nanochannels, thereby promoting correlated crystallographic orientations of BZL nanocrystals formed in neighboring channels. In practice, solidification of the interfacial bulk layer produces a polycrystalline BZL texture comprising crystallites extending over submillimeter-to-millimeter length scales. Such crystallites can extend across a large number of nanochannels and therefore act as common crystallographic templates for the confined BZL nanocrystals. The resulting spatial correlation of crystallographic orientations distinguishes this configuration from the uncorrelated crystallization occurring in free-standing membranes.

It should be noted that the substantial volumetric contraction of BZL upon crystallization may result in the formation of discontinuous crystalline regions, or solidified bridges, within the nanochannels, separated by nanoscale voids, as schematically illustrated in Figure~\ref{f1}(g,h). Similar contraction-induced void formation was previously observed in silica--BZL nanocomposites~\cite{Elkarout2023}. Consequently, BZL:AAO membranes exhibit pronounced scattering of both transmitted and reflected light. Such scattering becomes particularly pronounced when the porous substrate is not completely filled. Similar effects have been reported for the freezing of rare-gas solids and molecular crystals in nanoporous matrices~\cite{Soprunyuk2003,Wallacher2005a,Knorr2008}.

\subsection{Structural Characterization}

Raman spectroscopy was performed using a Renishaw inVia Reflex micro-Raman spectrometer with a spectral resolution of $\sim$1.6~cm$^{-1}$. A 532~nm laser was used as the radiation source, with a laser power of about 5~mW on the sample surface.

X-ray diffraction (XRD) was carried out on a DRON-4 diffractometer with Cu K$_{\alpha}$ radiation. Measurements were performed in the $\theta$--$2\theta$ geometry, with the AAO membrane oriented such that the X-ray beam probed the membrane surface, using a 1~cm long gap diaphragm, so that the illuminated sample area extended over about 1~cm in the scattering plane. Diffraction patterns were recorded in the $2\theta$ range of 10--50$^\circ$ with a step size of 0.01$^\circ$ and an integration time of 1~s per step.

\subsection{Second-Harmonic Generation (SHG) Measurements}

Because of the strong diffuse light scattering in the BZL:AAO nanocomposite samples, their nonlinear optical (NLO) properties were investigated by measuring the second-harmonic generation (SHG) response in both transmission and reflection geometries. The corresponding experimental setups are schematically illustrated in Figure~\ref{f1}, sections (j) and (l), respectively. Both setups are similar to those employed in our previous studies~\cite{Waszkowska2021,Elkarout2023,Shchur2024s}. A picosecond Nd:YAG laser ($\lambda = 1064$ nm, pulse energy $E_p = 50~\mu$J, pulse duration $\tau = 30$ ps, and repetition rate of 10 Hz) was used as the fundamental radiation source. The samples were mounted on a rotating optical stage, enabling measurements at different incident (azimuthal) angles $\alpha$. The polarization of the incident radiation was controlled using a Glan-Taylor prism and a rotatable half-wave plate. Four incident polarization directions were investigated, as defined in Figure~\ref{f1},sections (i) and (k): \textbf{S} corresponds to vertical polarization ($0^\circ$), \textbf{P} to horizontal polarization ($90^\circ$), \textbf{SP} to $45^\circ$, and \textbf{PS} to $135^\circ$.

In both setups, a long-focus lens, L$_1$, was used to adjust the incident laser power density to a nondestructive level, while a high-aperture, short-focus lens, L$_2$, collected the diffuse SHG radiation over a solid angle of 1.5 sr. Depending on the measurement geometry, the collected light was transmitted through or reflected from the sample. The SHG emission was then filtered using an interference filter F ($\lambda = 532$ nm) and detected by a photomultiplier (PM) coupled to a boxcar averager with an integration time of 0.1 s. A photodiode was used to monitor the energy of the fundamental laser pulses. For each incident polarization, azimuthal scans were performed by rotating the sample in $1^\circ$ increments and averaging the detected signal over 20 laser pulses at each angular position.

The NLO response of the BZL:AAO nanocomposites was characterized relative to reference samples measured under identical experimental conditions. For the transmission geometry, a 1mm-thick $\alpha$-quartz ($\alpha$-SiO$_2$) crystal plate was used in the $ooo$ configuration, corresponding to the $\chi_{111}$ tensor component. For the $90^\circ$ reflection geometry, crystalline $\alpha$-SiO$_2$ micropowder  was used as the reference naterials. The relative SHG response was defined as $R_{2\omega}=I_{2\omega}($BZL:AAO$)/I_{2\omega}(\alpha$-SiO$_2$), and this ratio was used throughout the study to characterize the SHG response of the nanocomposite samples.

\section*{Author Contributions}

\textbf{Conceptualization:} Patrick Huber, Bouchta Sahraoui, Andriy V. Kityk. \textbf{Methodology:} Andrii Z. Maxym, Junhui Lang, Bouchta Sahraoui, Olha Kityk, Monika Lelonek, Piotr Pawlik, Andriy V. Kityk. \textbf{Software:} Andrii Z. Maxym, Yaroslav Shchur. \textbf{Validation:} Olha Kityk. \textbf{Formal analysis:} Patrick Huber, Anatoliy S. Andrushchak, Yaroslav Shchur, Monika Lelonek, Piotr Pawlik, Jaros{\l}aw J{\k{e}}dryka, Andriy V. Kityk. \textbf{Investigation:} Andrii Z. Maxym, Junhui Lang, Yaroslav Shchur, Jaros{\l}aw J{\k{e}}dryka, Andriy V. Kityk. \textbf{Resources:} Olha Kityk, Robert Wielgosz, Monika Lelonek, Piotr Pawlik. \textbf{Data curation:} Andrii Z. Maxym, Junhui Lang, Yaroslav Shchur, Jaros{\l}aw J{\k{e}}dryka. \textbf{Writing -- original draft:} Patrick Huber, Bouchta Sahraoui, Andriy V. Kityk. \textbf{Writing -- review and editing:} Patrick Huber, Andriy V. Kityk. \textbf{Visualization:} Monika Lelonek, Piotr Pawlik. \textbf{Supervision:} Bouchta Sahraoui, Anatoliy S. Andrushchak, Robert Wielgosz. \textbf{Project administration:} Patrick Huber, Anatoliy S. Andrushchak, Robert Wielgosz. \textbf{Funding acquisition:} Patrick Huber, Bouchta Sahraoui, Anatoliy S. Andrushchak.

All authors have read and approved the final version of the manuscript.

\section*{Conflict of Interest}

There are no conflicts to declare.

\begin{acknowledgments}
The results presented are part of a project that has received funding from the European Union Horizon Europe research and innovation program under the Marie Sk{\l}odowska-Curie Grant agreement no.101086493. AK acknowledged the project cofinanced by the Polish Ministry of Education and Science under the program “cofinanced international projects”, project no. W26/HE/2023 (Dec. MEiN 5451/HE/2023/2). Funding by the Deutsche Forschungsgemeinschaft (DFG, German Research Foundation) within the collaborative research center CRC 986 “Tailor-Made Multi-Scale Materials Systems”, project number 192346071, and under Germany’s Excellence Strategy within the Cluster of Excellence BlueMat: Water-Driven Materials (EXC 3120/1 – 533771286), is also acknowledged. Generative AI tools were used in the preparation of this manuscript. Claude (Anthropic PBC) was used for language and stylistic refinement, for verifying bibliographic records against Crossref and for literature searching, and for drafting and revising passages of the Introduction and the Table of Contents entry. The Table of Contents graphic is a modified version of a layout proposed by ChatGPT (OpenAI). All AI-assisted content was checked against primary sources and edited by the authors, who rigorously reviewed the text and take full responsibility for the final publication.

\end{acknowledgments}

\section*{Data Availability Statement}
Data supporting the findings of this study are available from the corresponding author on a reasonable request.

\bibliographystyle{apsrev4-2}
\bibliography{references_KA}

\end{document}